\documentclass[11pt, a4paper]{article}

\usepackage{mathtools}
\usepackage{yhmath}
\usepackage{amsmath,bm,amssymb}
\usepackage{amsfonts}
\usepackage{bbold}
\usepackage{mathrsfs}
\usepackage{latexsym} 
\usepackage{cancel}
\usepackage{bm}
\usepackage{graphicx, rotating}
\usepackage{epstopdf}
\usepackage{epsfig}
\usepackage{latexsym}
\usepackage{color}
\usepackage[dvipsnames]{xcolor}
\usepackage{cite}
\usepackage{slashed}
\usepackage{hyperref}
\usepackage{comment}
\usepackage[utf8]{inputenc}
\usepackage{soul}
\usepackage{multirow}
\usepackage{multicol}
\usepackage{graphics,subfigure}
\usepackage[dvipsnames]{xcolor}
\usepackage{physics}
\usepackage{ulem}
\usepackage{dsfont}

\usepackage{tikz}
\usepackage{tikz-3dplot}
\usetikzlibrary{angles,quotes}

\tikzset{>=latex}

\hypersetup{colorlinks, citecolor=bluscuro, linkcolor=black, urlcolor=bluscuro}
\definecolor{rossos}{cmyk}{0,1,1,0.55}
\definecolor{bluscuro}{rgb}{0.15, 0.2, .85}
\definecolor{bluchiaro}{cmyk}{1,.3,0.,0.1}

\newcommand{\lp}{\left(}
\newcommand{\rp}{\right)}
\newcommand{\nn}{\nonumber}

\newcommand{\be}{\begin{equation}}
\newcommand{\ee}{\end{equation}}
\newcommand{\bea}{\begin{eqnarray}}
\newcommand{\eea}{\end{eqnarray}}

\newcommand\blfootnote[1]{%
  \begingroup
  \renewcommand\thefootnote{}\footnote{#1}%
  \addtocounter{footnote}{-1}%
  \endgroup
}

\def \cqfd {\hfill$\Box$}
\def\bea{\begin{eqnarray}}
\def\eea{\end{eqnarray}}

	\def \beq {\begin{equation}}
	\def \eeq {\end{equation}}
	\def \ba {\begin{array}}
	\def \ea {\end{array}}

	\def \ecart {\noalign{\medskip}}
	\def \dis {\displaystyle}

\def \RR {\mathbb R}

\newcommand{\orcid}{\includegraphics{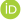}}
\newcommand{\orcidlink}[1]{\href{https://orcid.org/#1}{{\orcid}}}

\begin{document}

\begin{titlepage}
\begin{flushright}
IFT-UAM/CSIC-24-59
\end{flushright}

\begin{center} ~~\\
\vspace{0.5cm} 
\Large {\bf\Large Quantum tomography of helicity states \\ for general scattering processes.} 
\vspace*{1.5cm}

\normalsize{
{\bf 
Alexander Bernal
\orcidlink{0000-0003-3371-5320}
\blfootnote{
alexander.bernal@csic.es}} \\
 
\smallskip  \medskip
{\it Instituto de F\'\i sica Te\'orica, IFT-UAM/CSIC,}\\
\it{Universidad Aut\'onoma de Madrid, Cantoblanco, 28049 Madrid, Spain}}

\medskip

\vskip0.6in 

\end{center}

\centerline{ \large\bf Abstract }
\vspace{.5cm}

Quantum tomography has become an indispensable tool in order to compute the density matrix $\rho$ of quantum systems in Physics. Recently, it has further gained importance as a basic step to test entanglement and violation of Bell inequalities in High-Energy Particle Physics. In this work, we present the theoretical framework for reconstructing the helicity quantum initial state of a general scattering process. In particular, we perform an expansion of $\rho$ over the irreducible tensor operators $\{T^L_M\}$ and compute the corresponding coefficients uniquely by averaging, under properly chosen Wigner D-matrices weights, the angular distribution data of the final particles. Besides, we provide the explicit angular dependence of a novel generalisation of the production matrix $\Gamma$ and of the normalised differential cross section of the scattering. Finally, we re-derive all our previous results from a quantum-information perspective using the Weyl-Wigner-Moyal formalism and we obtain in addition simple analytical expressions for the Wigner $P$ and $Q$ symbols.

\vspace*{2mm}
\end{titlepage}

\tableofcontents

\section{Introduction}

The general description of a quantum mechanical system is encoded in a central mathematical entity known as {\it density matrix}, $\rho$. In this way, any physical accessible information of the system (any information attained by performing measurements over it) is theoretically obtained by computing expectation values of observables: $\langle {\cal O} \rangle=\Tr{{\cal O}\rho}$. Here, $\cal O$ stands for the Hermitian and linear operator associated with the considered experimental measurement. In particular, when dealing with systems with finite degrees of freedom, i.e. the associated Hilbert space has finite dimension, this quantum characterisation takes the form of a complex, unit-trace and Hermitian matrix. Therefore, in order to test with high precision either new or well-established models of Particle Physics, a practical and accurate reconstruction of $\rho$ is needed.

The procedure to determine this density matrix by performing measurements over the system is known in the quantum-information context as ``Quantum Tomography'' \cite{PhysRevLett.83.3103,PhysRevA.64.052312,PhysRevA.66.012303}. Special relevance has been given to the helicity density matrix since the works of E. Wigner\cite{wigner2012group}, M. Jacob \cite{Jacob:1958ply}, G. C. Wick \cite{Jacob:1959at} and J. Werle \cite{Werle1963127,Werle1963579,WERLE1963637} indicated a strong relation between angular distribution of final states and helicity amplitudes for general scattering processes. For instance, different methods for the tomography of single-particle systems \cite{Boudjema_2009,Aguilar_Saavedra_2016,Aguilar_Saavedra_2017} and of pairs of particles \cite{Bernreuther_2015,Afik_2021,Rahaman_2022,Ashby_Pickering_2023,PhysRevD.107.016012} have been proposed, all of them based on the reconstruction via the angular distribution data of the final particles. Nevertheless, they restrict themselves to final states reached after a chain of consecutive decays, while the plethora of other cases are not addressed. In this study we carry out the extension of the quantum tomography for all possible scattering processes. As a matter of fact, due to the transformation property under rotations of the irreducible tensor operators $\{T^L_M\}$\cite{doi:10.1142/0270,Aguilar_Saavedra_2016,Aguilar_Saavedra_2017,PhysRevD.107.016012} we base the reconstruction of $\rho$ on its expansion over this suitable basis, contrary to other approaches where either Cartesian vector and tensors \cite{Boudjema_2009,Martens:2017cvj,Rahaman_2022} or generalised Gell-Mann matrices \cite{Bertlmann_2008,Ashby_Pickering_2023} are used.

As aforementioned, once the density matrix is reconstructed several properties of the corresponding system can be tested. Namely, several theoretical papers have tackled the certification of entanglement as well as violation of Bell inequaities in pairs of top and anti-top \cite{Afik_2021,Fabbrichesi_2021,Severi_2022,Aguilar_Saavedra_2022,aguilarsaavedra2023postdecay,dong2023machine}, $\Lambda$ baryons \cite{Gong_2022}, $B^0\bar B^0$ pairs \cite{PhysRevD.104.056004}, charmonium decays \cite{Baranov:2008zzb,Chen:2013epa}, top-quarks\cite{Severi_2022,Aoude:2022imd,Fabbrichesi_2023,Afik_2023}, tau-leptons \cite{Fabbrichesi_2023,ma2023testing}, photons \cite{Fabbrichesi_2023}, electron-positron collisions \cite{Tornqvist:1980af}, positronium decays \cite{Acin:2000cs}, vector bosons \cite{BARR2022136866,Barr:2022wyq,PhysRevD.107.016012,bi2023new,Aguilar_Saavedra_2023,Bernal:2023ruk} and 2-2 scatterings \cite{PhysRevD.108.025015,Morales:2023gow}. On the experimental side, works concerning spin polarisation and spin correlations of top anti-top pairs \cite{ATLAS:2016bac,CMS:2019nrx,ATLAS:2022vym}, vector bosons \cite{ATLAS:2019bsc} as well as violation of Bell inequalities in ions \cite{article}, vector bosons \cite{Fabbri:2023ncz}, superconducting systems \cite{PMID:19779447}, nitrogen vacancies\cite{Pfaff_2012} and in photons \cite{Freedman:1972zza,Aspect:1982fx,PhysRevLett.89.240401} can also be found in the literature.

To sum up, the situation we will be referring along the work is a scattering process from $m$ initial particles to $n\geq2$ final particles. Actually, we are interested in the helicity density matrix of the initial state and the only information we have access to is the angular distribution of the final particles. From this starting point, we are able not only to perform the tomography of $\rho$ but also to give the explicit angular dependence of both the generalised production matrix $\Gamma$, see section \ref{sec:GenProdMat} for a rigorous definition, and of the normalised differential cross section of the process, the latter having been previously studied in \cite{Collins:1977iv,Lam:1980uc,Mirkes:1994eb,Mirkes:1994dp,Hagiwara:2006qe,Bern:2011ie,Stirling:2012zt}.

The paper is organised as follows: in section \ref{sec:StateRep} we present the quantum-mechanical formalism for the representation of many-particle systems used throughout the paper. Section \ref{sec:GenProdMat} is devoted to the construction of the generalised production matrix $\Gamma$, while in section \ref{sec:DensMat} we develop the quantum tomography of the density matrix $\rho$ from its relation with both the normalised differential cross section and the generalised production matrix, elaborating on the case in which $\Gamma$ factorises. We illustrate the quantum tomography method during section \ref{sec:Examples} by providing simple High-Energy Particle Physics examples. In section \ref{sec:WignerPQ} we relate the theoretical framework here introduced to the Weyl-Wigner-Moyal formalism \cite{Li_2013}. Finally, in section \ref{sec:Conclusions} we summarise the whole work and present our conclusions. Appendix \ref{sec:CasesProdMat} collects the explicit form of the generalised production matrix for different scattering processes in terms of the initial and final particle's number, whereas Appendix \ref{sec:SigmaCond} is dedicated to mathematical technicalities omitted in the main text.

\newpage

\section{State representation for relativistic many-particle systems}\label{sec:StateRep}

This section is devoted to summarise and recall the formalism of state representations for relativistic many-particle systems developed in references \cite{Werle1963127,Werle1963579,WERLE1963637}. We emphasise on the indispensable concepts for the work.

Let us consider an arbitrary system of $n$ particles with definite $3$-momenta $\vec{p}_i$, spins $s_i$ and helicities $\lambda_i\in\{-s_i,\dots,s_i\}$, such that the total linear momentum $\vec{\chi}=\sum_{i=1}^n \vec{p_i}=\vec{0}$. In particular, we will denote this centre-of-mass spatial reference frame by $\mathcal{RF}$ and consider $\vec{p}_i=p_i\, \hat{p}_i$,  with $p_i$ and $\hat{p}_i$ the $3$-momentum modulus and unit vector of particle $i$.  

\begin{figure}[ht]
    \centering
    \includegraphics[height=6.5cm]{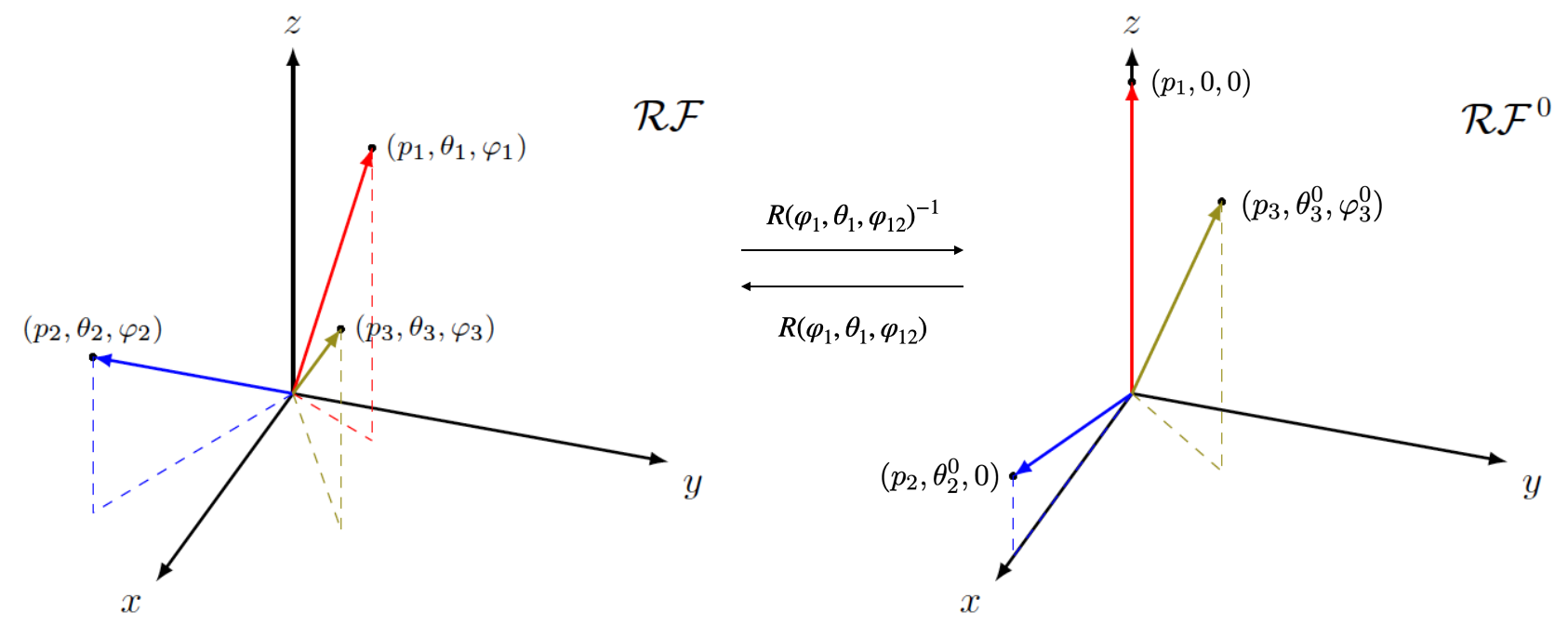}
    \caption{On the left, a sketch of the most general centre-of-mass spatial reference frame of $n$ particles $\mathcal{RF}$ (only three of the $n$ particles are displayed for clarity). On the right, a sketch of the special centre-of-mass reference frame of $n$ particles $\mathcal{RF}^{\,0}$ fixed by particles 1 and 2 (only three of the $n$ particles are displayed). The corresponding Euler rotation to move from one frame to another is also shown.}
    \label{Fig:RefFrame}
\end{figure}

Based on these $3$-momenta, we can fix a particularly useful centre-of-mass reference frame as displayed in Fig.\ref{Fig:RefFrame}, which we denote by $\mathcal{RF}^{\,0}$. Namely, we set the $3$-momentum of one of the particles (from now on particle $1$) to define the $z$ axis via $\hat{z}=\hat{p}_1$, while the $3$-momentum of a second particle (from now on particle $2$) is considered to lie in the $x>0$ half of the $xz$ plane, so that $\hat{y}=\frac{\hat{p}_1\cross\hat{p}_2}{|\hat{p}_1\cross\hat{p}_2|}$. Therefore, $\mathcal{RF}^{\,0}$ is completely characterised by the $3$-momenta of the two selected particles, which we have assumed to be distinguishable from the rest and between them. 

We denote the $3$-momenta of particle $i$ in the $\mathcal{RF}^{\,0}$ frame by $\vec{p}_i^{\ 0}$. For instance, the spherical coordinates (modulus as well as polar and azimuthal angles) characterising $\vec{p}_i^{\ 0}$ take the values 
\begin{equation} \label{SphCoord0}
    \lp p_1, \theta_{1}^{\,0},\varphi_{1}^{\,0}\rp=\lp p_1,0,0\rp, \quad \lp p_2, \theta_{2}^{\,0},\varphi_{2}^{\,0}\rp=\lp p_2, \theta_{2}^{\,0},0 \rp,\quad \lp p_i, \theta_{i}^{\,0},\varphi_{i}^{\,0}\rp \hbox{ for } i=3,\dots, n .
\end{equation}
We notice that only $3n-3$ non-trivial coordinates are needed to describe the system in this frame. On the other hand, the corresponding $3n$ coordinates of $\vec{p}_i$ in a general $\mathcal{RF}$ are obtained from the ones in $\mathcal{RF}^{\,0}$ by performing an Euler rotation $R\lp \alpha,\beta,\gamma\rp$ of the whole system as a rigid body. 

The inverse operation can also be done: the $\mathcal{RF}^{\,0}$ description is recovered by applying a $R\lp\alpha,\beta,\gamma\rp^{-1}$ rotation to the system described in $\mathcal{RF}$, see Fig.\ref{Fig:RefFrame}. In this sense, the physical meaning of the Euler angles characterising the rotation is easily obtained once particles $1$ and $2$ are identified in $\mathcal{RF}$. Specifically, $\lp\alpha,\beta\rp=\lp \varphi_1,\theta_1\rp$ are the angular components of particle $1$ in $\mathcal{RF}$  whereas $\gamma=\varphi_{12}$ is the azimuthal coordinate of particle $2$ after applying an initial $R(\alpha,\beta,0)^{-1}$ rotation to the system, i.e. after setting the $3$-momentum of particle $1$ to be in the $z$ axis.

Following this convention, the quantum state of a general $n$-particle system with definite $3$-momenta and helicities as well as null total linear momentum is given by
\begin{equation}\label{RepPi}
\prod_{i=1}^n\ket {\vec{p}_i\, \lambda_i}=\hat{D}\lp R\rp\prod_{i=1}^n\ket {\vec{p}_i^{\ 0}\, \lambda_i}=\ket{R\, p^0 \,\lambda},
\end{equation}
here $\hat{D}\lp R\rp$ is the unitary representation of the rotation acting on the quantum states, $R=R\lp \alpha,\beta,\gamma\rp$ (for simplicity we omit the angle dependence of $R$) and $\vec{p}_i=R\, \vec{p}_i^{\ 0}$. In addition, we have denoted by $\lambda$ the whole set of particle's helicities and by $p^0$ the set of $3n -3$ spherical coordinates needed to describe the system in $\mathcal{RF}^{\,0}$, as specified in Eq.\eqref{SphCoord0}. The orthogonality conditions for these states read
\begin{eqnarray}
    &\hspace{-0.8cm}\bra{R'\, p^{0\, '}\,\lambda'}\ket{R\,p^0\,\lambda}=\delta\lp R,R' \rp\delta\lp p^0\, p^{0\, '};\lambda\lambda'\rp, \hbox{ with } \delta\lp R,R' \rp=\delta\lp\alpha-\alpha'\rp\delta\lp\cos{\beta}-\cos{\beta}'\rp\delta\lp\gamma-\gamma'\rp
    \\ \ecart \nn
    &\dis{\hbox{ and } \delta\lp p^0\, p^{0\, '};\lambda\lambda'\rp=\frac{\delta\lp p_1-p_1'\rp}{p_1^2}\,\frac{\delta\lp p_2-p_2'\rp}{p_2^2}\,\frac{\delta\lp\theta_2^{\,0}-\theta_2^{\,0\,'}\rp}{\sin{\theta_2^{\,0}}}\,\prod_{i=3}^n\delta_3\lp\vec{p}_i^{\ 0}-\vec{p}_i^{\ 0\,'}\rp\,\prod_{j=1}^n\delta_{\lambda_j\,\lambda_j'}.
    }
\end{eqnarray}

Following \cite{WERLE1963637}, a more convenient representation of \eqref{RepPi} is possible. Namely, the $p^0$ set of quantum numbers is replaced by the total energy of the system $E$, the total null linear momentum $\vec{\chi}=\vec{0}$ and a set $\kappa$ of $3n-7$ parameters, functions of $p^0$, to be chosen depending on the case of interest. Whenever $3n-7\leq0$ we take $\kappa=\varnothing$, the empty set. Besides, due to the conservation of the total energy and since we are dealing with center-of-mass frames, both $E$ and $\vec{\chi}$ will play no role in what follows. Hence, the state \eqref{RepPi} becomes $\ket{R\, \kappa\, \lambda}$, whose orthogonality conditions are
\begin{eqnarray}\label{RepKappa}
    &\dis{\bra{R'\,  \kappa'\, \lambda'}\ket{R\, \kappa\, \lambda}=\delta\lp R',R \rp\delta\lp\kappa\,\kappa';\lambda\lambda'\rp
    }\\ \ecart \nn
    &\dis{ \hbox{ and } \delta\lp\kappa\,\kappa';\lambda\lambda'\rp=\prod_{i=1}^{3n-7}\delta \lp\kappa_i-\kappa_i'\rp\,\prod_{j=1}^n\delta_{\lambda_j\,\lambda_j'}.
    }
\end{eqnarray}
Finally, let us also introduce the relation between the states in Eq.\eqref{RepKappa} and those of definite angular momentum:
\begin{eqnarray} \label{JtoR}
    \ket{J\, M\, \Lambda\, \kappa\, \lambda}=\frac{\lp J+1/2\rp^{1/2}}{2\pi}\int dR\, D^J_{M\,\Lambda}\lp R\rp^* \ket{R\, \kappa\, \lambda}, \hbox{ with }\\ \ecart \nn
    \bra{R\, \kappa'\, \lambda'}\ket{J\, M\, \Lambda\, \kappa\, \lambda}=\frac{\lp J+1/2\rp^{1/2}}{2\pi}D^J_{M\,\Lambda}\lp R\rp^*\,\delta\lp\kappa\,\kappa';\lambda\lambda'\rp.
\end{eqnarray}
The function $D^J_{M\,M'}\lp R\rp \delta_{J\,J'}=\bra{J\, M}\hat{D}\lp R\rp\ket{J'\, M'}$ is the Wigner D-matrix associated with the Euler rotation $R$, $dR$ is the differential with respect to the angles characterising $R$ and the quantum number $\Lambda$ corresponds to the projection of the total angular momentum of the system over $\hat{p}_1$. From the orthogonality conditions of the Wigner D-matrices,
\begin{equation}\label{OrthCondD}
    \int dR\, D^{J'}_{M'\,\Lambda'}\lp R\rp^* \, D^J_{M\,\Lambda}\lp R\rp=\frac{8\pi^2}{2J+1}\ \delta_{J\,J'}\ \delta_{M\,M'}\ \delta_{\Lambda\,\Lambda'},
\end{equation}
we get 
\begin{equation}
    \bra{J'\, M'\, \Lambda'\, \kappa'\, \lambda'}\ket{J\, M\,\Lambda\, \kappa\, \lambda}=\delta_{J\,J'}\delta_{M\,M'}\delta_{\Lambda\,\Lambda'}\delta\lp\kappa\,\kappa';\lambda\lambda'\rp.
\end{equation}
The inverse relation concerning the states in Eq.\eqref{JtoR} is 
\begin{equation}\label{RtoJ}
    \ket{R\, \kappa\, \lambda}=\sum_J \sum_{M\,\Lambda}\frac{(J+1/2)^{1/2}}{2\pi}D^J_{M\,\Lambda}(R)\ket{J\, M\, \Lambda\, \kappa\, \lambda}.
\end{equation}

A further final comment is needed: during this section we have assumed that the particles we are dealing with have a fixed and known mass $m_i$. Nevertheless this may not be the case, for example when working with off-shell particles. For those scenarios, we also need to treat $m_i$ as an extra quantum number per particle (to add in the parameter set $\kappa$) in order to have a complete representation of the system. Specific examples are analysed in Appendix \ref{sec:CasesProdMat}.

\section{Generalised production matrix}\label{sec:GenProdMat}
In this section we generalise the concept of production (or decay) matrix to any kind of scattering process. This matrix was previously defined e.g. in \cite{haber1994spin,Boudjema_2009} in terms of the helicity amplitudes, but only accounting for decay and 2 to 2 scattering processes. In addition, in order to give a practical characterisation of the matrix we also introduce a generalisation for any scattering process of the so-called {\it reduced helicity amplitudes} \cite{haber1994spin,Boudjema_2009,Rahaman_2022}, which were defined as well only in the context of decay and 2 to 2 scattering processes. For instance, we define the \textit{generalised production matrix} by 
\begin{equation}
\Gamma_{\bar{\lambda}\, \bar{\lambda}'}\propto\sum_\lambda {\cal M}_{\lambda\,\bar{\lambda}}{\cal M}_{\lambda\,\bar{\lambda}'}^*
\end{equation}
where the normalisation $\Tr{\Gamma}=1$ is left to impose. Here, as done in \cite{Werle1963127,Werle1963579,WERLE1963637}, we have denoted the initial state quantities with an overbar and the final state ones without it. The amplitudes ${\cal M}_{\lambda\,\bar{\lambda}}$ are the so-called helicity amplitudes, whose expression in terms of the scattering matrix $S$ and the states introduced in Eq.\eqref{RtoJ} is
\begin{equation}
    {\cal M}_{\lambda\,\bar{\lambda}}=\bra{R\, \kappa\, \lambda}S\ket{\bar{R}\,\bar{\kappa}\,\bar{\lambda}}.
\end{equation}
The rotations $\bar{R},\,R$ are performed with respect to a fixed reference frame, $\mathcal{RF}^{\,0}$, and along with $\lp\bar \kappa,  \bar\lambda\rp$ as well as $\lp\kappa,\lambda\rp$ completely characterise the initial and final states. Furthermore, by definition of \eqref{RepPi} and \eqref{RtoJ}, the property 
\begin{equation}
    \ket{R\, \kappa\, \lambda}=\hat{D}\lp R\rp\ket{\mathbb{1}\, \kappa\, \lambda}
    \label{StateProp}
\end{equation}
holds, where the notation $\mathbb{1}$ refers to the absence of a rotation and $\hat{D}\lp R\rp$ is the unitary representation of the rotation $R$ in the Hilbert space of interest.

As will be justified later on, see Eq.\eqref{MasterEq} in section \ref{sec:DensMat} and Eq.\eqref{GammaPOVM} in section \ref{sec:WignerPQ}, the relevant matrix for our final goal is the transposed of the generalised production matrix, $\Gamma^T$. Therefore, using the expression of ${\cal M}_{\lambda\,\bar{\lambda}}$:
\begin{eqnarray}
    \Gamma^T_{\bar{\lambda}\, \bar{\lambda}'}&\propto&\sum_\lambda \bra{\bar{R}\,\bar{\kappa}\,\bar{\lambda}}S^\dagger\ket{R\, \kappa\, \lambda} \bra{R\, \kappa\, \lambda}S\ket{\bar{R}\,\bar{\kappa}\,\bar{\lambda}'}\\ \ecart
    &=&  \bra{\bar{R}\,\bar{\kappa}\,\bar{\lambda}}\left[\sum_\lambda \lp S^\dagger \ket{R\, \kappa\, \lambda} \bra{R\, \kappa\, \lambda}S\rp \right]\ket{\bar{R}\,\bar{\kappa}\,\bar{\lambda}'}\nn.
\end{eqnarray}
Using property \eqref{StateProp} of $\ket{R\, \kappa\, \lambda}$ as well as the rotation invariance of the scattering matrix $S$, we get
\begin{eqnarray}
    \Gamma^T_{\bar{\lambda}\, \bar{\lambda}'}&\propto&\bra{\bar{R}\,\bar{\kappa}\,\bar{\lambda}} \hat{D}\lp R\rp \left[\sum_\lambda\lp S^\dagger\ket{\mathbb{1}\, \kappa\, \lambda} \bra{\mathbb{1}\, \kappa\, \lambda}S \rp \right]\hat{D}\lp R\rp^{-1} \ket{\bar{R}\,\bar{\kappa}\,\bar{\lambda}'} \\ \ecart \nn
    &=&\bra{\bar{\mathbb{1}}\,\bar{\kappa}\,\bar{\lambda}}\lp \hat{D}\lp \bar{R}\rp^{-1} \hat{D}\lp R\rp\rp \left[\sum_\lambda\lp S^\dagger\ket{\mathbb{1}\, \kappa\, \lambda} \bra{\mathbb{1}\, \kappa\, \lambda}S \rp \right]\lp \hat{D}\lp \bar{R}\rp^{-1} \hat{D}\lp R\rp\rp^{-1} \ket{\bar{\mathbb{1}}\,\bar{\kappa}\,\bar{\lambda}'}\\ \ecart \nn
    &=&\bra{\bar{\mathbb{1}}\,\bar{\kappa}\,\bar{\lambda}}\hat{D}\lp \bar{R}^{-1} R \rp \left[\sum_\lambda\lp S^\dagger\ket{\mathbb{1}\, \kappa\, \lambda} \bra{\mathbb{1}\, \kappa\, \lambda}S \rp \right]\hat{D}\lp \bar{R}^{-1} R \rp^{-1} \ket{\bar{\mathbb{1}}\,\bar{\kappa}\,\bar{\lambda}'}.
\end{eqnarray}
This last equation implies that in the helicity basis for the initial states, $\{\ket{\bar{\mathbb{1}}\,\bar{\kappa}\,\bar{\lambda}}\}_{\bar{\lambda}}$, the matrix $\Gamma^T$ only depends on the relative rotation between the initial and final states, i.e. $\Gamma^T\lp \bar{R},R\rp=\Gamma^T\lp\bar{R}^{-1} R\rp$, and it is given by rotating $\Gamma^T\lp\bar{\mathbb{1}},\mathbb{1}\rp=\Gamma^T\lp\mathbb{1}\rp$ accordingly, as it should happen for a generic quantum-mechanical operator in a rotated system \cite{doi:10.1142/0270}:
\begin{equation}
    \Gamma^T\lp \bar{R},R\rp=\Gamma^T\lp \bar{R}^{-1} R\rp=\hat{D}\lp \bar{R}^{-1} R \rp\Gamma^T(\mathbb{1})\hat{D}\lp \bar{R}^{-1} R \rp^{-1}.
\end{equation}
Furthermore, it is worth mentioning that instead of considering two different rotations $\lp\bar{R},\,R\rp$ with respect to the fixed reference frame $\mathcal{RF}^{\,0}$, we can always set the initial state of the scattering process to define $\mathcal{RF}^{\,0}$. Thus, $\bar{R}=\bar{\mathbb{1}}$ and the whole relative rotation only comes from a rotation of the final state with respect to this reference frame, $\bar{R}^{-1} R=R$. Following the physical interpretation given in section \ref{sec:StateRep}, the set of Euler angles characterising $R$ are denoted by $\lp \alpha, \beta,\gamma\rp=\lp \varphi_1, \theta_1,\varphi_{12}\rp=\Omega$. In consequence,
\begin{equation}
    \bar{R}^{-1} R=R= R \lp \varphi_1, \theta_1,\varphi_{12}\rp= R \lp \Omega \rp\implies \Gamma^T\lp \bar{R}^{-1}R\rp=\Gamma^T\lp R\rp=\hat{D}\lp R \rp\Gamma^T(\mathbb{1})\hat{D}\lp R \rp^{-1}.
\end{equation}
From now on, we will consider $\Gamma^T\equiv\Gamma^T\lp\mathbb{1}\rp$ and specify the argument $\lp\mathbb{1}\rp$ only when needed.

The exact expression for $\Gamma^T$ is presented in full detail in Appendix \ref{sec:CasesProdMat}. For instance, some relevant features of this matrix are discussed for the following scattering processes (labelled in terms of the initial and final particle's number):
\begin{equation*} 
    1)\ \bar{1}\to 2 \quad\quad  2)\ \bar{1}\to n \quad\quad  3)\ \bar{2}\to 2 \quad\quad   4)\ \bar{2}\to n \quad\quad 5)\ \bar{m}\to 2 \quad\quad  6)\ \bar{m}\to n.
\end{equation*}
In general, the generalised production matrix can be expanded in the canonical basis $\{e_{\sigma_i\,\sigma_i'}\}$ via a sum over the possible spin projections $\sigma_i\in\{-s_i,\dots,s_i\}$ of the initial particles:
\begin{eqnarray}
    &\dis{\Gamma^T \propto 
    \sum_{\sigma\, \sigma'} 
    a_{\sigma\, \sigma'}\, e_{\sigma\, \sigma'}, \hbox{ with } \sigma= \lp \sigma_1,\dots,\sigma_m\rp
    } \hbox{ and same with primes,}\\ \ecart \nn
    &\dis{\Gamma^T_{\sigma\, \sigma'}\propto a_{\sigma\, \sigma'}=\sum_\lambda\bra{\bar{\mathbb{1}}\,\bar{\kappa}\,\sigma} S^\dagger\ket{\mathbb{1}\, \kappa\, \lambda} \bra{\mathbb{1}\, \kappa\, \lambda}S \ket{\bar{\mathbb{1}}\,\bar{\kappa}\,\sigma'},\quad e_{\sigma\, \sigma'}=\bigotimes_{i=1}^m e_{\sigma_i\, \sigma_i'},\quad [e_{\sigma_i\, \sigma'_i}]_{\bar{\lambda}_i,\bar{\lambda}_i'}=\delta_{\sigma_i,\bar{\lambda}_i}\ \delta_{\sigma_i',\bar{\lambda}_i'}.
    }
\end{eqnarray}
I.e., $e_{\sigma_i\, \sigma_i '}$ is a unit matrix with the $\lp \sigma_i, 
\sigma_i'\rp$-element equal to $1$ and all the others null. In addition, the $\lp \sigma, 
\sigma\rp$-elements $a_{\sigma\, \sigma'}=a_{\sigma\, \sigma'}\lp\bar{\kappa},\kappa\rp$ are a generalisation of the so-called {\it reduced helicity amplitudes} \cite{haber1994spin,Boudjema_2009,Rahaman_2022} and encode the kinematics not included in $\Omega$. Nonetheless, during the current and next sections the $\lp\bar{\kappa},\kappa\rp$ dependence will be in general omitted as it is not relevant.

Moreover, imposing $\Tr{\Gamma^T}=\Tr{\Gamma}=1$ one can deduce the normalisation factor in $\Gamma^T$ as a function of $a_{\sigma\, \sigma'}$:
\begin{equation}
    \Tr{e_{\sigma\, \sigma'}}=\delta_{\sigma\, \sigma'}\Longrightarrow \Tr{\sum_{\sigma\, \sigma'} a_{\sigma\, \sigma'} e_{\sigma\, \sigma'}}=\sum_{\sigma} a_{\sigma\, \sigma}, \hbox{ where } a_{\sigma\, \sigma}=\sum_\lambda \left|\bra{\mathbb{1}\, \kappa\, \lambda}S \ket{\bar{\mathbb{1}}\,\bar{\kappa}\,\sigma}\right|^2\in\mathbb{R}^{+}.
\end{equation}
Hence, defining $a_{+}\equiv\sum_{\sigma} a_{\sigma\, \sigma}\in\mathbb{R}^{+}$,
\begin{equation}
    \Gamma^T=\frac{1}{a_{+}}\sum_{\sigma\, \sigma'} a_{\sigma\, \sigma'} e_{\sigma\, \sigma'} \Longrightarrow \Gamma^T\lp R\rp=\frac{1}{a_{+}}\sum_{\sigma\, \sigma'} a_{\sigma\, \sigma'} \hat{D}\lp R \rp e_{\sigma\, \sigma'} \hat{D}\lp R \rp^{-1}.
\end{equation}

\section{Reconstruction of the density matrix}\label{sec:DensMat}

The goal of this section is to develop the quantum tomography for the initial state helicity density matrix $\rho$. The main idea is to expand $\rho$ over the basis of irreducible tensor operators $\{T^L_M\}$ and then compute each coefficient in the decomposition by performing an integration over the angular distribution of the final state. Alternative proposals for a quantum tomography method in this context \cite{Boudjema_2009,Rahaman_2022,Ashby_Pickering_2023} are valid only for decay processes and use  either the Cartesian vectors or the generalised Gell-Mann matrices instead of the irreducible tensor operators as a basis for the decomposition of the density matrix.

To accomplish the final result, we first 
deepen in some properties of $\Gamma^T\lp R\rp$, providing its decomposition over $\{T^L_M\}$ and the Wigner D-matrices $\{D^L_{M\,\Lambda}\lp R\rp\}$. This decomposition will make explicit the $\Gamma^T\lp R\rp$ transformation properties under rotations of the system as well as its kinematic dependence.

Secondly, given the relation between the production matrix, the normalised differential cross section and the density matrix in helicity space of the initial state \cite{haber1994spin,Boudjema_2009,Rahaman_2022}:
\begin{equation}
    \dfrac{1}{\sigma}\dfrac{d \sigma}{d\Omega\, d \bar{\kappa}\, d \kappa}={\cal N}\Tr{\rho\, \Gamma^T\lp R\rp}
    \label{MasterEq}
\end{equation}
(with $\cal N$ a normalisation constant to be fixed later on) we give the, as far as we know, first decomposition of the normalised differential cross section over $\{D^L_{M\,\Lambda}\lp R\rp\}$ and the coefficients of the expansion of $\rho$ under the irreducible tensor operators.

Finally, we provide a practical and experimentally viable way of reconstructing the density matrix of any system (to use e.g. in further studies about spin correlation or entanglement properties) once the normalised differential cross section of the scattering process is measured.

As has just been anticipated, in order to perform the tomography of $\rho$ we need to expand both $\Gamma^T\lp R\rp$ and $\rho$ over the convenient basis of irreducible tensor operators, whose transformation under rotations will be crucial for the incoming reasoning. Let us briefly explain the properties of this basis. We follow the notation presented in \cite{doi:10.1142/0270}, where the irreducible tensor operators are $\left\{T_M^L\lp s\rp\right\}_{L,\, M}$. In the context we are working on, $s=\lp s_1,\dots,s_m\rp$ represents the vector of spins of the initial particles, which furthermore fixes the dimension of $T^L_M\lp s\rp$ to be equal to $d\cross d$, with $d= \prod_{i=1}^m d_i= \prod_{i=1}^m \lp 2 s_i +1\rp$. 

In addition, $L\in\{0,\dots, 2 s_T\}$ and $M\in\{-L,\dots,L\}$, where $s_T$ is seen as an ``effective spin'' corresponding to the whole system, i.e. $d= 2 s_T+1$. Denoting by $\sigma_T,\sigma_T'\in\{-s_T,\dots,s_T\}$ the possible projections of $s_T$, an explicit expression for the elements of $T^L_M(s)$ is \cite{doi:10.1142/0270}
\begin{equation}
    \left[T^L_M\lp s\rp\right]_{\sigma_T\, \sigma_T'}=\lp 2L +1\rp ^{1/2} C^{s_T\, \sigma_T}_{s_T\, \sigma_T'\, L\, M},
\end{equation}
where $C^{j\, m}_{j_1\, m_1\, j_2\, m_2}=\bra{j_1 m_1 j_2 m_2}\ket{j_1 j_2 j m}$ are the Clebsch-Gordan coefficients, which following the Condon and Shortley convention are chosen to be real. Throughout the paper, we will denote $T^L_M\equiv T^L_M\lp s\rp$ when there is no ambiguity and recover the $s$-vector notation otherwise.

Finally, some other properties concerning the Hermitian conjugation as well as the orthogonality relations among the tensors are:
\begin{equation}
   \lp T^L_M\rp^\dagger=\lp-1\rp^M T^L_{-M},\quad\Tr{T^L_M\, \lp T^{L'}_{M'}\rp^\dagger}=\Tr{T^L_M\, \lp T^{L'}_{M'}\rp^T}=d\, \delta_{L\,L'}\delta_{M\,M'}.
\end{equation}
The latter one leads to a theoretical simple way of computing the coefficient of a general operator ${\cal O}$ with respect to this basis:
\begin{equation}
    {\cal O}=\frac{1}{d}\sum_{L\,M} {\cal O}_{L\,M} T^L_M, \hbox{ with } {\cal O}_{L\,M}=\Tr{{\cal O}\, \lp T^L_M\rp^\dagger}.
    \label{OpExpansion}
\end{equation}

We will apply the last result to first obtain the expansion of $\Gamma^T\lp R\rp$ and then $\rho$, getting the desired quantum tomography. We start by determining the coefficients in the expansion of the matrix $e_{\sigma\, \sigma'}$:
\begin{equation}
    \Tr{e_{\sigma\, \sigma'}\, \lp T^L_M\rp^\dagger}= \Tr{e_{\sigma\, \sigma'}\, \lp T^L_M\rp^T}=\left[T^L_M\right]_{\sigma_T\, \sigma_T'}=\lp 2L +1\rp ^{1/2} C^{s_T\, \sigma_T}_{s_T\, \sigma_T'\, L\, M},
\end{equation}
where $\sigma_T$ is defined via the scalar product of $\sigma$ with a suitable vector only dependent on the dimensions of the Hilbert spaces:
\begin{eqnarray}
    &\sigma_T=\sigma\cdot d^{(v)} \hbox{ and } \sigma_T'=\sigma'\cdot d^{(v)}, \hbox{ with } d^{(v)}=(d^{(v)}_1,\dots,d^{(v)}_m),\\ \ecart
    &d^{(v)}_i=\prod_{j=i+1}^m d_j, \hbox{ for } i<m \hbox{ and } d^{(v)}_m=1.
\end{eqnarray}
In particular, the vector $d^{(v)}$ can be seen as a mathematical tool connecting the spin projection $\sigma_i$ of every particle with the spin projection $\sigma_T$ of the ``effective spin'' $s_T$ corresponding to the whole system. 
Moreover, using the properties of the Clebsch-Gordan coefficients:
\begin{equation}
    \Tr{e_{\sigma\, \sigma'}\ \lp T^L_M\rp^\dagger}= \lp 2L +1\rp ^{1/2} C^{s_T\, \sigma_T}_{s_T\, \sigma_T'\, L\, M}=\lp 2L +1\rp ^{1/2} C^{s_T\, \sigma_T}_{s_T\, \sigma_T'\, L\, \sigma_T^{-}} \delta_{M\, \sigma_T^{-}},
\end{equation}
with $\sigma_T^{-}=\sigma_T-\sigma_T'=\lp\sigma-\sigma'\rp\cdot d^{(v)}$. Finally
\begin{equation}
    e_{\sigma\, \sigma'}=\frac{1}{d}\sum_{L\,M}\left[\lp 2L +1\rp ^{1/2} C^{s_T\, \sigma_T}_{s_T\, \sigma_T'\, L\, \sigma_T^{-}}\ \delta_{M\, \sigma_T^{-}}\right]T^L_M=\frac{1}{d}\sum_{L}C_{L\,\sigma\,\sigma'} T^L_{\sigma_T^{-}},
\end{equation}
where $C_{L\,\sigma\,\sigma'}\equiv\lp 2L +1\rp ^{1/2} C^{s_T\, \sigma_T}_{s_T\, \sigma_T'\, L\, \sigma_T^{-}}\in\RR$. Plugging the expansion into  $\Gamma^T$ and rearranging the sum in $\lp\sigma, \sigma'\rp$:
\begin{eqnarray}
    \Gamma^T&=&\frac{1}{a_{+}}\sum_{\sigma\, \sigma'} a_{\sigma\, \sigma'} \left[\frac{1}{d}\sum_{L} C_{L\,\sigma\,\sigma'} T^L_{\sigma_T^{-}} \right]=\frac{1}{d}\sum_{L}\sum_{\sigma_T^{-}}\Bigg[\dfrac{1}{a_{+}}\sum_{\substack{\sigma\, \sigma'\\ \lp\sigma-\sigma'\rp\cdot d^{(v)}=\sigma_T^{-}}}a_{\sigma\, \sigma'}\, C_{L\,\sigma\,\sigma'}\Bigg]T^L_{\sigma_T^{-}}\\ \ecart \nn
    &=&\frac{1}{d}\sum_{L\,\sigma_T^{-}}\tilde{B}_{L\,\sigma_T^{-}}T^L_{\sigma_T^{-}}. 
\end{eqnarray}
The newly introduced coefficient is defined as
\begin{equation}
    \tilde{B}_{L\,\sigma_T^{-}}\lp \bar \kappa,\kappa\rp\equiv\dfrac{1}{a_{+}\lp \bar \kappa,\kappa\rp}\sum_{\substack{\sigma\, \sigma'\\ \lp\sigma-\sigma'\rp\cdot d^{(v)}=\sigma_T^{-}}}a_{\sigma\, \sigma'}\lp \bar \kappa,\kappa\rp\, C_{L\,\sigma\,\sigma'}
    \label{BTildDef}
\end{equation}
and carries the dependence on $\lp \bar \kappa,\kappa\rp$ via the reduced helicity amplitudes $a_{\sigma\, \sigma'}\lp \bar \kappa,\kappa\rp$. 

Furthermore, making use of the Hermitian-conjugation properties of $\Gamma^T$, $e_{\sigma\, \sigma'}$ and $T^L_M$, we can deduce the ones for $a_{\sigma\, \sigma'},\,\tilde{B}_{L\,\sigma_T^{-}}$ and $C_{L\,\sigma\,\sigma'}$. Namely, it is easy to see that
\begin{equation}
    a_{\sigma\, \sigma'}^*=a_{\sigma'\, \sigma},\quad \tilde{B}_{L\,\sigma_T^{-}}^*=\lp-1\rp^{\sigma_T^{-}}\tilde{B}_{L\,-\sigma_T^{-}}, \quad C_{L\,\sigma'\,\sigma}=\lp-1\rp^{\sigma_T^{-}}C_{L\,\sigma\,\sigma'}.
\end{equation}
For the special $\sigma_T^{-}=0$ case, the condition $\lp\sigma-\sigma'\rp\cdot d^{(v)}=0$ is equivalent to $\sigma=\sigma'$. The proof of this statement is presented in Appendix \ref{sec:SigmaCond}, and the simplified expressions for the coefficients in that case read 
\begin{eqnarray}\label{BTild0}
     &\dis{\left.a_{\sigma\, \sigma'}\right|_{\sigma_T^{-}=0}=a_{\sigma\, \sigma}\in \RR^{+},\quad \left.C_{L\,\sigma\,\sigma'}\right|_{\sigma_T^{-}=0}=\lp 2L +1\rp ^{1/2} C^{s_T\, \sigma_T}_{s_T\, \sigma_T\, L\, 0},
     }\\ \ecart\nn
     &\dis{\tilde{B}_{L}\equiv\tilde{B}_{L\,0}=\dfrac{1}{a_{+}}\sum_{\substack{\sigma\, \sigma'\\ \sigma=\sigma'}}a_{\sigma\, \sigma'}\, C_{L\,\sigma\,\sigma'}=\frac{\lp 2L +1\rp^{1/2}}{a_+}\sum_{\sigma}a_{\sigma\, \sigma}\ C^{s_T\, \sigma_T}_{s_T\, \sigma_T\, L\, 0}\in \RR.
     }
\end{eqnarray}

Coming back to the computation of $\Gamma^T\lp R\rp$, as explained in section \ref{sec:GenProdMat} we obtain it by rotating $\Gamma^T$ with the unitary representation of $R$, which we recall refers to the Euler rotation connecting the initial and final states in the scattering. Denoting $\sigma_T^{-}=M'$ for simplicity in what comes, 
\begin{equation}
    \Gamma^T\lp R\rp= \hat{D}\lp R \rp \Gamma^T \hat{D}\lp R \rp^{-1} =\frac{1}{d}\sum_{L\,M'}\tilde{B}_{L\,M'}\lp \hat{D}\lp R \rp T^L_{M'} \hat{D}\lp R \rp^{-1}\rp.
\end{equation}
Taking into account the transformation properties under rotations of the irreducible tensor operators, 
\begin{eqnarray}\nn\label{GammaDecom}
    &\dis{\hat{D}\lp R \rp T^L_{M'} \hat{D}\lp R \rp^{-1}=\sum_M D^L_{M\,M'}\lp R\rp T^L_M\implies
    }\\ \ecart 
    &\dis{\implies\boxed{\Gamma^T\lp R\rp=\frac{1}{d}\sum_{L\,M}\left[\sum_{M'}\tilde{B}_{L\,M'}D^L_{M\,M'}\lp R\rp\right]T^L_M.}
    }
\end{eqnarray}
with $D^L_{M\,M'}\lp R \rp$ the Wigner D-matrix associated with the rotation $R$. Expression \eqref{GammaDecom} gives the exact decomposition of the generalised production matrix in terms of irreducible tensor operators. It is worth mentioning that the kinematic dependence on the coefficients has been factorised as
\begin{eqnarray}
\tilde{B}_{L\,M'}=\tilde{B}_{L\,M'}\lp\bar{\kappa}, \kappa \rp,\quad D^L_{M\,M'}\lp R\rp=D^L_{M\,M'}\lp\Omega\rp.
\end{eqnarray}
The $\rho$ matrix is also expanded with respect to the irreducible tensor operators following the recipe in Eq.\eqref{OpExpansion}. Hence,
\begin{equation}\label{rhoExp}
      \rho=\frac{1}{d}\sum_{L\,M} A_{L\,M}\ T^L_M,\hbox{ with } A_{L\,M}=A_{L\,M}\lp \bar{\kappa}\rp.
\end{equation}
The coefficients $A_{L\,M}$ do not depend on the parameters $\lp\kappa,R\rp$ characterising the final state, as $\rho$ is a mathematical entity only referred to the initial one.

In order to experimentally reconstruct $\rho$, let us recall Eq.\eqref{MasterEq}, which relates $\rho$ to both the normalised differential cross section and the production matrix. The normalisation constant $\cal N$ appearing in Eq.\eqref{MasterEq} is such that $\dis{\int d\Omega\,d \bar{\kappa}\, d\kappa\, \frac{1}{\sigma}\frac{d \sigma}{d\Omega\, d \bar{\kappa}\, d \kappa}=1}$, therefore
\begin{equation}
    {\cal N}^{-1}=\int d\Omega\,d \bar{\kappa}\, d\kappa \Tr{\rho\, \Gamma^T\lp R\rp}=\int d \bar{\kappa}\, d\kappa \Tr{\rho\,\int d\Omega\, \Gamma^T\lp R\rp}.
\end{equation}
In the last step we have used that the only $\Omega$-dependent quantity is $\Gamma^T\lp R\rp$. Actually, the integration over $\Omega$ can be performed using the decomposition in Eq.\eqref{GammaDecom} as well as property \eqref{OrthCondD}:
\begin{equation}
    \int d\Omega\, \Gamma^T\lp R\rp=\frac{1}{d}\sum_{L\,M}\sum_{M'}\tilde{B}_{L\,M'}\left[\int d\Omega\,D^L_{M\,M'}\lp R\rp\right]T^L_M=\frac{8\pi^2}{d}\tilde{B}_{00}\,T^0_0=\frac{8\pi^2}{d}\mathbb{1}.
\end{equation}
In the end, since $\rho$ is normalised we get
\begin{equation}
    {\cal N}^{-1}=\frac{8\pi^2}{d}\int d \bar{\kappa}\, d\kappa \Tr{\rho}=\dfrac{8\pi^2 \bar{K} K}{d},\quad  \hbox { with } \bar K=\int d\bar\kappa\,\hbox{ and }\,  K=\int d\kappa.
\end{equation}
Using in Eq.\eqref{MasterEq} this value of $\cal N$ along with Eq.\eqref{rhoExp} gives
\begin{equation}
     \dfrac{1}{\sigma}\dfrac{d \sigma}{d\Omega\, d \bar{\kappa}\, d \kappa}=\dfrac{d}{8\pi^2  \bar{K} K}\left[\frac{1}{d}\sum_{L\,M} A_{L\,M} \Tr{T^L_M\, \Gamma^T\lp R\rp}\right]=\dfrac{1}{8\pi^2  \bar{K} K}\sum_{L\,M} A_{L\,M} \Tr{T^L_M\, \Gamma^T\lp R\rp}.
\end{equation}
Last term in the previous equation can be obtained from the expansion and Hermiticity of $\Gamma^T\lp R\rp$ as well as general properties of the trace:
\begin{eqnarray}
    \Tr{T^L_M\, \Gamma^T\lp R\rp}&=&\dis{\Tr{\lp T^L_M\rp^T\, \Gamma\lp R\rp}=\Tr{\lp T^L_M \rp^\dagger\, \Gamma^T\lp R\rp}^*
    }\\ \ecart \nn
    &=&\dis{\left[\sum_{M'}\tilde{B}_{L\,M'} D^L_{M\,M'}\lp R\rp\right]^*=\sum_{M'}\tilde{B}_{L\,M'}^* D^L_{M\,M'}\lp R\rp^*}.
\end{eqnarray}
Plugging the above result in the expression of the normalised differential cross section leads to its expansion in terms of both the Wigner D-matrices and the coefficients $\tilde{B}_{L\,M'}$,
\begin{equation}\label{CrosSectionExpans}
    \boxed{\dfrac{1}{\sigma}\dfrac{d \sigma}{d\Omega\, d \bar{\kappa}\, d \kappa}=\dfrac{1}{8\pi^2 \bar{K} K}\sum_{L\,M} A_{L\,M} \sum_{M'}\tilde{B}_{L\,M'}^* D^L_{M\,M'}\lp R\rp^*.}
\end{equation}
The presence of the coefficients $A_{L\, M}$ in the right-hand side of the latter equation is clear. In order to isolate them we use the orthogonality conditions for the Wigner D-matrices introduced in Eq.\eqref{OrthCondD}. Finally, making explicit the dependence on $\lp\bar{\kappa}, \kappa \rp$:
\begin{eqnarray}
    &\dis{\int d\Omega\, \left[\dfrac{1}{\sigma}\dfrac{d \sigma}{d\Omega\, d \bar{\kappa}\, d \kappa}\right]\lp\dfrac{2L+1}{4\pi}\rp^{1/2}D^L_{M\,M'}\lp \Omega\rp=
    }\\ \ecart \nn
    &\dis{=\lp\dfrac{2L+1}{4\pi}\rp^{1/2}\dfrac{1}{8\pi^2 \bar{K} K}\sum_{\hat{L}\,\hat{M}} A_{\hat{L}\,\hat{M}}\lp\bar{\kappa}\rp \sum_{\hat{M}'} \tilde{B}_{\hat{L}\,\hat{M}'}\lp\bar{\kappa}, \kappa \rp^* \int d\Omega\, D^{\hat{L}}_{\hat{M}\,\hat{M}'}\lp \Omega\rp^*\,D^L_{M\,M'}\lp \Omega\rp}\\ \ecart \nn
    &\dis{=\dfrac{1}{\left[4\pi \lp 2L+1\rp\right]^{1/2}}\left[\frac{\tilde{B}_{L\,M'}\lp\bar{\kappa}, \kappa \rp^*}{\bar{K} K} \right]A_{L\,M}\lp\bar{\kappa}\rp=\frac{B_{L\,M'}\lp\bar{\kappa}, \kappa \rp^*}{4\pi}A_{L\,M}\lp\bar{\kappa}\rp}.
\end{eqnarray}
Here, we have defined the coefficient
\begin{equation}
    B_{L\,M'}\lp\bar{\kappa}, \kappa \rp\equiv\lp\frac{4\pi}{2L+1}\rp^{1/2} \frac{\tilde{B}_{L\,M'}\lp\kappa, \bar{\kappa}\rp}{\bar{K} K}
\end{equation}
which shares the same conjugation relations than $\tilde{B}_{L,M'}\lp\bar{\kappa}, \kappa \rp$.

In this sense, the main goal of the work is achieved and the expansion of $\rho$ in the $\left\{T_M^L\right\}_{L,\, M}$ basis can be fully reconstructed via the integration of the normalised differential cross section under a properly chosen kernel,
\begin{equation}\label{QTrho}
    \boxed{A_{L\,M}\lp\bar{\kappa}\rp=\dfrac{4\pi}{ B_{L\,M'}\lp\bar{\kappa}, \kappa \rp^*}\int d\Omega\, \left[\dfrac{1}{\sigma}\dfrac{d \sigma}{d\Omega\, d \bar{\kappa}\, d \kappa}\right]\lp\frac{2L+1}{4\pi}\rp^{1/2}D^L_{M\,M'}\lp \Omega\rp.}
\end{equation}
This expression generalises a first approach given for a specific decay process in \cite{PhysRevD.107.016012}. The procedure for extracting a density matrix coefficient for fixed $L$ and $M$ stands as long as $\tilde{B}_{L\,M'}\lp\bar{\kappa}, \kappa \rp\neq0$ for at least one $M'$. The $B_{L\,M'}\lp\bar{\kappa}, \kappa \rp=0\ \forall M'$ scenario would imply, by Eq.\eqref{CrosSectionExpans}, that the normalised differential cross section has no component associated with the angular momentum $L$. Thus, it is not that the method is incomplete in order to extract the density matrix, but that the scattering process considered carries no such angular information, so it is physically impossible to reconstruct that part of the initial state from the process in hand. 

Special relevance has the $M'=0$ case, for which
\begin{eqnarray}\label{QTrhoY}
    &\dis{\int d\Omega\,\left[\dfrac{1}{\sigma}\dfrac{d \sigma}{d\Omega\, d \bar{\kappa}\, d \kappa}\right]\lp\frac{2L+1}{4\pi}\rp^{1/2}D^L_{M\,0}\lp \Omega\rp=\frac{B_{L\,0}\lp\bar{\kappa}, \kappa \rp^*}{4\pi}A_{L\,M}\lp\bar{\kappa}\rp
    }\\ \ecart \nn
    &\dis{\iff\int d\Omega\, \left[\dfrac{1}{\sigma}\dfrac{d \sigma}{d\Omega\, d \bar{\kappa}\, d \kappa}\right]Y^M_L\lp\theta, \varphi\rp^*=\frac{B_L\lp\bar{\kappa}, \kappa \rp}{4\pi}A_{L\,M}\lp\bar{\kappa}\rp,
    }
\end{eqnarray}
where we have used the relation between the Wigner D-matrices and the spherical harmonics as well as the definition of a new coefficient $B_L$ for consistency with the notation used in \cite{PhysRevD.107.016012}:
\begin{eqnarray}\label{B0}
    &\dis{D^L_{M\,0}\lp R\rp^*=D^L_{M\,0}\lp \varphi, \theta, \varphi_{12} \rp^*=\lp\frac{4\pi}{2L+1}\rp^{1/2} Y^M_L\lp\theta, \varphi\rp,
    }\\ \ecart \nn
    &\dis{B_L\lp\bar{\kappa},\kappa\rp\equiv B_{L\,0}\lp\bar{\kappa},\kappa\rp=\lp\frac{4\pi}{2L+1}\rp^{1/2} \frac{\tilde{B}_{L}\lp\bar{\kappa}, \kappa \rp}{\bar{K} K}=\frac{\sqrt{4\pi}}{a_+\lp\bar{\kappa},\kappa\rp\bar{K} K}\sum_{\sigma}\,a_{\sigma\, \sigma}\lp\bar{\kappa},\kappa\rp C^{s_T\, \sigma_T}_{s_T\, \sigma_T\, L\, 0} \in\RR.
    }
\end{eqnarray}
In general, when $B_L\lp\bar{\kappa},\kappa\rp\neq0$, $M'=0$ constitutes the simplest choice in order to apply the quantum tomography procedure here developed.

\subsection{Factorisation of the production matrix}\label{sec:FacScat}

Among all of the scattering processes taking place in Particle Physics, there are several of them for which the $\Gamma$ matrix factorises. Some examples have been already studied in the literature \cite{Boudjema_2009,PhysRevD.107.016012,Ashby_Pickering_2023,Rahaman_2022}. In this section we explain how the quantum tomography of the density matrix is simplified in those scenarios.

In particular, let us consider a scattering process of the form
\begin{equation} \hspace{-0.1cm}
    \lp \bar{A}_1\, \bar{B}_1\, \bar{C}_1\dots\rp\,\lp \bar{A}_2\, \bar{B}_2\, \bar{C}_2\dots\rp\dots\,\lp \bar{A}_N\, \bar{B}_N\, \bar{C}_N\dots\rp\to \lp A_1\, B_1\, C_1\dots\rp\,\lp A_2\, B_2\, C_2\dots\rp\dots\,\lp A_N\, B_N\, C_N\dots\rp, 
\end{equation}
where we have the factorisation $\lp \bar{A}_j\, \bar{B}_j\, \bar{C}_j\dots\rp\to\lp A_j\, B_j\, C_j\dots\rp\, \forall j$. Therefore, the production matrix of the whole process also factorises as 
\begin{equation}
    \Gamma=\bigotimes_{j=1}^N\Gamma_j\lp R_j\rp,
\end{equation}
with $\Gamma_j\lp R_j\rp$ being the production matrix of each individual $j$ process presented above and $R_j$ the associated Euler rotation. This factorisation leaves an impact in Eq.\eqref{MasterEq}, which is rewritten as
\begin{equation}
    \dfrac{1}{\sigma}\dfrac{d \sigma}{d\Omega\, d \bar{\kappa}\, d \kappa}={\cal N}\Tr{\rho\, \lp\bigotimes_{j=1}^N\Gamma^T_j\lp R_j\rp\rp}.
\end{equation}
Here, $\lp\Omega,\bar{\kappa},\kappa\rp$ denotes the \textit{vector} of kinematic parameters $\lp\Omega_j,\bar{\kappa}_j,\kappa_j\rp$ for each process.

In this context, instead of expanding our density matrix in the $\left\{T_M^L\lp s\rp\right\}_{L,\, M}$ basis, we choose the preferred $\left\{\bigotimes_{j=1}^N T_{M_j}^{L_j}\lp s_j\rp\right\}_{L_j,\, M_j}$ one, with $\left\{T_{M_j}^{L_j}\lp s_j\rp\right\}_{L_j,\, M_j}$ being the basis of irreducible tensor operators to be used for the individual $\lp \bar{A}_j\, \bar{B}_j\, \bar{C}_j\dots\rp\to\lp A_j\, B_j\, C_j\dots\rp$ process. Thus, the density matrix expansion reads
\begin{equation}
    \rho=\frac{1}{d}\sum_{L_1\,L_2\,\dots L_N}\sum_{M_1\,M_2\,\dots M_N} A_{L_1\,M_1,\,L_2\,M_2,\dots,\,L_N\,M_N}\ \bigotimes_{j=1}^N T_{M_j}^{L_j}\lp s_j\rp.
\end{equation}
Applying a similar reasoning than the one for the general case, 
\begin{equation}\label{QTrhoProd}\hspace{-0.5cm}
\boxed{A_{L_1\,M_1,\,L_2\,M_2,\dots,\,L_N\,M_N}\lp\bar{\kappa}\rp=\dfrac{(4\pi)^N}{ \prod_{j=1}^N B_{L_j\,M_j'}\lp\bar{\kappa}_j, \kappa_j \rp^*}\int d\Omega \left[\dfrac{1}{\sigma}\dfrac{d \sigma}{d\Omega\, d \bar{\kappa}\, d \kappa}\right]\left[\prod_{j=1}^N\lp\frac{2L_j+1}{4\pi}\rp^{1/2}D^{L_j}_{M_j\,M_j'}\lp \Omega_j\rp\right].}     
\end{equation}
Finally, let us elaborate on the situation for which all the $j$ processes considered correspond to particle's decays, this is $N=m$ and
\begin{equation}
    \bar{A}_1\,\bar{A}_2\dots\bar{A}_m\to \lp A_1\, B_1\, C_1\dots\rp\,\lp A_2\, B_2\, C_2\dots\rp\dots\,\lp A_m\, B_m\, C_m\dots\rp.
\end{equation}
In these examples, the $A_{L_1\,M_1,\,L_2\,M_2,\dots,\,L_m\,M_m}$ coefficients have a clear physical interpretation. They are related to the spin polarisations and the spin correlations between the particles taking part in the process:
\begin{itemize}
    \item $L_j=L_{j_0}\delta_{j\,j_0} \longrightarrow A_{0\,0,\dots,L_{j_0}\,M_{j_0},\dots,0\,0}$ coincides with the spin polarisation vector of the particle $j_0$.
    \item $L_j=0$ except for $L_{j_1},L_{j_2} \longrightarrow A_{0\,0,\dots,L_{j_1}\,M_{j_1},\dots,L_{j_2}\,M_{j_2},\dots,0\,0}$ represents the spin correlation matrix of particles $j_1$ and $j_2$.
    $$\vdots$$
    \item $L_j\neq0 \quad \forall j \longrightarrow A_{L_1\,M_1,\,L_2\,M_2,\dots,\,L_m\,M_m}$ gives the spin correlation tensor of the whole system.
\end{itemize}

\section{High-Energy Particle Physics examples}\label{sec:Examples}
 In this section we provide several simple examples in the context of High-Energy Physics (all of them verifying $\bar \kappa=\kappa=\varnothing$) for which the quantum tomography method here developed applies. With these processes we aim to exemplify how to apply our method to different initial states, to show the simplicity it provides compared to other proposals \cite{Boudjema_2009,Ashby_Pickering_2023} and to point out the new avenues it opens because of covering a broader class of scattering processes.
 
We begin with the quantum tomography of single-particle states, then we move to analyse different two-particle systems and we end with what, to the best of our knowledge, is the
first detailed theoretical example given in the literature for the quantum tomography of a 3-particle system. 
 
\subsection{$t\to b W^+\to bl^+\nu_l$}\label{sec:Example1}
The reconstruction of the density matrix $\rho$ for the top quark constitutes one of initial steps to perform further analysis either of single-particle systems \cite{Boudjema_2009,Ashby_Pickering_2023} or composed ones \cite{Severi_2022,Aoude:2022imd,Fabbrichesi_2023,Afik_2023}. Since the top quark is a spin-$1/2$ particle, we have $s=1/2$ and the irreducible tensor operators entering in the decomposition of $\rho$ are those with $L=0,1$ and $M=-L,\dots,L$:
\begin{equation}
    \rho=\dfrac{1}{2}\sum_{L=0}^1\sum_{M=-L}^L A_{L\,M}\, T^L_M(1/2).
\end{equation}
The explicit expressions for $T^L_M(1/2)$ are:
\begin{equation*}
   T^0_0(1/2)= \mathbb{1}_2,\quad T^1_1(1/2)=
   -\sqrt{2}\left(\begin{array}{cc}
         0&  1\\
         0  & 0
    \end{array}\right),\quad T^1_0(1/2)=
   \left(\begin{array}{cc}
         1& 0 \\
         0  & -1
    \end{array}\right),\quad T^1_{-1}(1/2)=-T^1_1(1/2)^T.
\end{equation*}
Besides, the production matrix for a spin-$1/2$ particle is given by \cite{Boudjema_2009}
\begin{equation}\label{eq:Gamma1/2}
   \Gamma^T(\theta,\varphi)=
   \left(\begin{array}{cc}
         \dfrac{1+\alpha\cos{\theta}}{2}& \dfrac{\alpha\sin{\theta}}{2} e^{-i \varphi} \\ \ecart
         \dfrac{\alpha\sin{\theta}}{2} e^{i \varphi}   & \dfrac{1-\alpha\cos{\theta}}{2}
    \end{array}\right),
\end{equation}
where $\alpha$ is known as the spin analysing power of the decay products. Its expression is given in terms of the reduced helicity amplitudes, which for decay processes (see Appendix \ref{sec:CasesProdMat12} for further details) fulfil $a_{\sigma\,\sigma'}=a_{\sigma\,\sigma} \delta_{\sigma\, \sigma'}$. Namely, denoting  $a_\sigma=a_{\sigma\,\sigma}$ we have
\begin{equation}
    \alpha=\frac{a_{1/2}-a_{-1/2}}{a_{1/2}+a_{-1/2}}.
\end{equation}
In particular, $\alpha_b\simeq-0.41$ when considering the bottom quark as particle 1, i.e. the one whose momentum defines $(\theta,\varphi)$ \cite{Brandenburg:2002xr}. On the contrary and when neglecting the final leptons masses, the decay $W^+\to l^+\nu_l$ joined to the posterior measure of the charged lepton momentum direction act, due to maximal parity violation, as a projective measure over the helicity of the $W^+$. Therefore, we can take the charged lepton as particle 1 and its corresponding spin analysing power is $\alpha_{l^+}\simeq 1$.  

The trace $\Tr{T^L_M\, \Gamma^T\lp \theta,\varphi\rp}$ lets us identify the coefficients $\tilde B_{L}$ and $B_{L}$, defined in Eq.\eqref{BTild0} and Eq.\eqref{B0} respectively, which are necessary to perform the tomography:
\begin{equation}\label{BCoef1/2}
\begin{array}{ccc}
     \Tr{T^0_0(1/2)\, \Gamma^T\lp \theta,\varphi\rp}=D^0_{0\,0}\lp\Omega\rp^*&\implies& \tilde B_{0}=1,\quad B_{0}=\sqrt{4\pi},  \\ \ecart
     \Tr{T^1_M(1/2)\, \Gamma^T\lp \theta,\varphi\rp}=\alpha D^1_{M\,0}\lp\Omega\rp^*&\implies& \tilde B_{1}=\alpha,\quad B_{1}=\sqrt{\dfrac{4\pi}{3}}\alpha .
\end{array}
\end{equation}
Plugging these results in Eq.\eqref{QTrho} leads to:
\begin{equation}
A_{0\,0}=1,\quad A_{1\,M}=\dfrac{3}{\alpha}\int d\Omega\, \left[\dfrac{1}{\sigma}\dfrac{d \sigma}{d\Omega}\right]D^1_{M\,0}\lp \Omega\rp.
\end{equation}
As expected $A_{0\,0}=1$ due to normalisation of the density matrix. Regarding $A_{1\,M}$, these coefficients give the components of the spin polarisation vector of the top quark \cite{ATLAS:2022vym,Bernreuther_2015} and are easily obtained by integrating the product of $D^1_{M\,0}\lp \Omega\rp$ times the experimentally determined normalised differential cross section. 

\subsection{$Z\to l^+ l^-$}\label{sec:Example2}
In the same fashion as with the top quark, the density matrix for single vector bosons has been studied both from the theoretical \cite{Aguilar_Saavedra_2016,Aguilar_Saavedra_2017,Boudjema_2009} and experimental \cite{ATLAS:2019bsc} side. This process corresponds to the decay of a spin-$1$ particle, hence $s=1$ and $L=0,1,2$:
\begin{equation}
    \rho=\dfrac{1}{3}\sum_{L=0}^2\sum_{M=-L}^L A_{L\,M}\, T^L_M(1).
\end{equation}
The relevant irreducible tensor operators are given by:
\begin{equation*}
    T^0_0(1)=\mathbb{1}_3,\quad T^1_1(1)=\sqrt{\frac{3}{2}}
    \lp
    \begin{array}{ccc}
     0 & -1 & 0 \\
     0 & 0 & -1  \\
     0 & 0 & 0 \\
    \end{array}
    \rp,\quad 
    T^1_0(1)=\sqrt{\frac{3}{2}}
    \lp
    \begin{array}{ccc}
     1 & 0 & 0 \\
     0 & 0 & 0  \\
     0 & 0 & -1 \\
    \end{array}
    \rp, 
\end{equation*}
with $T^1_{-1}(1)=-T^1_{1}(1)^T$ and
\begin{equation*}
T^2_2(1)=\sqrt{3}
\left(
\begin{array}{ccc}
 0 & 0 & 1 \\
 0 & 0 & 0  \\
 0 & 0 & 0 \\
\end{array}
\right),\quad
T^2_{1}(1)=\sqrt{\frac{3}{2}}
\left(
\begin{array}{ccc}
 0 & -1 & 0 \\
 0 & 0 & 1  \\
 0 & 0 & 0 \\
 \end{array}
\right),\quad
T^2_0(1)=\frac{1}{\sqrt{2}}
\left(
\begin{array}{ccc}
 1 & 0 & 0 \\
 0 & -2 & 0  \\
 0 & 0 & 1 \\
\end{array}
\right),
\end{equation*}
with $T^2_{-2}(1)=T^2_{2}(1)^T$ and $T^2_{-1}(1)=-T^2_{1}(1)^T$. Moreover, the production matrix for a spin-$1$ particle is given by \cite{Boudjema_2009}:
{\footnotesize
\begin{equation}\label{eq:Gamma1}
   \Gamma^T(\theta,\varphi)=
   \dfrac{1}{4}\left(
\begin{array}{ccc}
 1+\delta+(1-3\delta)\cos^2\theta+2\alpha\cos\theta & \sqrt{2}\sin\theta(\alpha+(1-3\delta)\cos\theta)e^{-i\varphi}& (1-3\delta)(1-\cos^2\theta) e^{-i2\varphi}\\ \ecart
 \sqrt{2}\sin\theta(\alpha+(1-3\delta)\cos\theta)e^{i\varphi}& 4\delta+2(1-3\delta)\sin^2\theta& \sqrt{2}\sin\theta(\alpha-(1-3\delta)\cos\theta)e^{-i\varphi} \\ \ecart
 (1-3\delta)(1-\cos^2\theta) e^{i2\varphi} & \sqrt{2}\sin\theta(\alpha-(1-3\delta)\cos\theta)e^{i\varphi} & 1+\delta+(1-3\delta)\cos^2\theta-2\alpha\cos\theta
\end{array}
\right),
\end{equation}}
with $(\theta,\varphi)$ the angles defining the momentum direction of one of the final charged leptons. In this case, 2 parameters are needed to characterise the $\Gamma$ matrix. Their expressions in terms of the diagonal reduced helicity amplitudes, $a_{\sigma}=a_{\sigma\,\sigma}$, are:
\begin{equation}
    \alpha=\frac{a_{1}-a_{-1}}{a_{1}+a_{-1}+a_0},\quad \delta=\frac{a_{0}}{a_{1}+a_{-1}+a_0}.
\end{equation}
Since we are taking the $Z$ boson as the initial vector boson, one can neglect the masses of the final leptons (in comparison to $m_Z$) and therefore obtain $\delta=0$ and $\alpha\simeq-0.13$. Proceeding as in last subsection, we derive the coefficients $\tilde B_{L}$ and $B_{L}$ by computing traces over $\Gamma^T(\theta,\varphi)$:
\begin{equation}\label{BCoef1}
\begin{array}{ccc}
     \Tr{T^0_0(1)\, \Gamma^T\lp \theta,\varphi\rp}=D^0_{0\,0}\lp\Omega\rp^*&\implies& \tilde B_{0}=1,\quad B_{0}=\sqrt{4\pi},  \\ \ecart
     \Tr{T^1_M(1)\, \Gamma^T\lp \theta,\varphi\rp}=\sqrt{\dfrac{3}{2}}\alpha D^1_{M\,0}\lp\Omega\rp^*&\implies& \tilde B_{1}=\sqrt{\dfrac{3}{2}}\alpha,\quad B_{1}=\sqrt{2\pi}\alpha,  \\ \ecart
     \Tr{T^2_M(1)\, \Gamma^T\lp \theta,\varphi\rp}=\dfrac{(1-3\delta)}{\sqrt{2}}D^2_{M\,0}\lp\Omega\rp^*&\implies& \tilde B_{2}=\dfrac{(1-3\delta)}{\sqrt{2}},\quad B_{2}=\sqrt{\dfrac{2\pi}{5}}(1-3\delta).
\end{array}
\end{equation}
Plugging these values in Eq.\eqref{QTrho} gives:
\begin{equation}
A_{0\,0}=1,\quad A_{1\,M}=\dfrac{\sqrt{6}}{\alpha}\int d\Omega\, \left[\dfrac{1}{\sigma}\dfrac{d \sigma}{d\Omega}\right]D^1_{M\,0}\lp \Omega\rp,\quad A_{2\,M}=\dfrac{5\sqrt{2}}{1-3\delta}\int d\Omega\, \left[\dfrac{1}{\sigma}\dfrac{d \sigma}{d\Omega}\right]D^2_{M\,0}\lp \Omega\rp.
\end{equation}
We get $A_{0\,0}=1$ due to normalisation of the density matrix. Regarding the coefficients defining the spin polarisation vector \cite{ATLAS:2019bsc}, $A_{1\,M}$ and $A_{2\,M}$, they are obtained by integrating the product of either $D^1_{M\,0}\lp \Omega\rp$ or $D^2_{M\,0}\lp \Omega\rp$ times the experimentally determined normalised differential cross section.

\subsection{$t\bar t\to (bl^+\nu_l) (\bar b l^-\bar \nu_l) $}\label{sec:Example3}
This is the first process involving an initial 2-particle state. Namely, we are dealing with the simplest bipartite initial state (2 qubits), which is composed by a $t\bar t$ pair. Several properties have been studied concerning this system, ranging from spin correlations \cite{ATLAS:2016bac,CMS:2019nrx} to entanglement and Bell inequality violations \cite{Afik_2021,Fabbrichesi_2021,Severi_2022,Aguilar_Saavedra_2022,aguilarsaavedra2023postdecay,dong2023machine}.

The quantum state decomposition for this 2-spin-$1/2$ particles, $s_1=s_2=1/2$, reads
\begin{equation}
    \rho=\frac{1}{4}\sum_{L_1\,L_2=0}^1\sum_{M_1\,M_2} A_{L_1\,M_1,\,L_2\,M_2}\,T_{M_1}^{L_1}\lp 1/2\rp \otimes T_{M_2}^{L_2}\lp 1/2\rp.
\end{equation}
As detailed in section \ref{sec:FacScat}, the coefficients $A_{L_1\,M_1,\,0\,0}$ ($A_{\,0\,0,\,L_2\,M_2}$) constitute the spin polarisation of particle 1 (2), while $A_{L_1\,M_1,\,L_2\,M_2}$ with $L_1,L_2\neq0$ gives the spin correlation matrix of the 2-fermion system.
Furthermore, this is the simplest scenario in which the production matrix factorises:
\begin{equation}
    \Gamma^T(\theta_1,\varphi_1,\theta_2,\varphi_2)=\Gamma_1^T(\theta_1,\varphi_1)\otimes\Gamma_2^T(\theta_2,\varphi_2).
\end{equation}
Here $\Gamma^T_i(\theta_i,\varphi_i)$ is the production matrix associated with the decay of fermion $i$ and the angles $(\theta_i,\varphi_i)$ refer to the momentum direction of one of the final particles in its corresponding decay. Their explicit expressions are given in Eq.\eqref{eq:Gamma1/2}.

The components of the now 2-dimensional vectors $\tilde B_{L}=(\tilde B_{L_1},\tilde B_{L_2})$ and $B_{L}=(B_{L_1}, B_{L_2})$ coincide with the coefficients $\tilde B_{L}$ and $B_{L}$ computed in the section \ref{sec:Example1}. In addition, for each decay we could have in principle different spin analysers $\alpha$, so
\begin{equation}
\begin{array}{cc}
   \tilde B_{0}=1,  & \tilde B_{L_i=1}=\alpha_i,  \\ \ecart
    B_{0}=\sqrt{4\pi}, & B_{L_i=1}=\sqrt{\dfrac{4\pi}{3}}\alpha_i.
\end{array}
\end{equation}
Replacing their values in Eq.\eqref{QTrhoProd}:
\begin{equation}
\begin{array}{c}
\dis A_{1\,M_1,\,0\,0}=\dfrac{3}{\alpha_1}\int d\Omega_1 d\Omega_2\, \left[\dfrac{1}{\sigma}\dfrac{d \sigma}{d\Omega_1 d\Omega_2}\right]D^1_{M_1\,0}\lp \Omega_1\rp, \bigskip \\ 
A_{\,0\,0,\,1\,M_2}=\dfrac{3}{\alpha_2}\int d\Omega_1 d\Omega_2\, \left[\dfrac{1}{\sigma}\dfrac{d \sigma}{d\Omega_1 d\Omega_2}\right]D^1_{M_2\,0}\lp \Omega_2\rp, \bigskip \\ 
\dis A_{1\,M_1,\,1\,M_2}=\dfrac{9}{\alpha_1\alpha_2}\int d\Omega_1 d\Omega_2\, \left[\dfrac{1}{\sigma}\dfrac{d \sigma}{d\Omega_1 d\Omega_2}\right]D^1_{M_1\,0}\lp \Omega_1\rp\,D^1_{M_2\,0}\lp \Omega_2\rp.
\end{array}
\end{equation}
and $A_{0\,0,\,0\,0}=1$ due to normalisation. Finally, the values of $\alpha_1$ and $\alpha_2$ will depend on which final particle (either the bottom quark or the charged lepton) is defining $(\theta_i,\varphi_i)$ and their explicit values coincide with those given in section \ref{sec:Example1}, $\alpha_b\simeq-0.41$ and $\alpha_l\simeq1$.

\subsection{$V_1 V_2\to (f_1\bar f_1) (f_2\bar f_2) $}\label{sec:Example4}
In this scattering process we are dealing with another scenario in which the production matrix factorises. For instance, we start with two vector bosons ($s_1=s_2=1$) decaying each into two final fermions. Exhaustive theoretical and experimental studies of these kinds of systems have been carried out \cite{Aguilar_Saavedra_2023,Bernal:2023ruk,Barr:2022wyq,BARR2022136866,bi2023new,ATLAS:2019bsc}, so we provide a practical method for extracting its density matrix following an initial idea shown in \cite{PhysRevD.107.016012}. The complete production matrix is given by:
\begin{equation}
\Gamma^T(\theta_1,\varphi_1,\theta_2,\varphi_2)=\Gamma_1^T(\theta_1,\varphi_1)\otimes\Gamma_2^T(\theta_2,\varphi_2).
\end{equation}
Here $\Gamma^T_i(\theta_i,\varphi_i)$ is the production matrix associated with the decay of boson $i$ and the angles $(\theta_i,\varphi_i)$ refer to the momentum direction of one of the final fermions in its corresponding decay. Their explicit expressions are given in Eq.\eqref{eq:Gamma1}. With respect to the density matrix, its decomposition is given by:
\begin{equation}
    \rho=\frac{1}{9}\sum_{L_1\,L_2=0}^2\sum_{M_1\,M_2} A_{L_1\,M_1,\,L_2\,M_2}\,T_{M_1}^{L_1}\lp 1\rp \otimes T_{M_2}^{L_2}\lp 1\rp.
\end{equation}
The coefficients $A_{L_1\,M_1,\,0\,0}$ ($A_{\,0\,0,\,L_2\,M_2}$) are interpreted as the elements of the spin polarisation vector of particle 1 (2), while $A_{L_1\,M_1,\,L_2\,M_2}$ with $L_1,L_2\neq0$ gives the spin correlation matrix of the 2-boson system.

The components of the vectors $\tilde B_{L}=(\tilde B_{L_1},\tilde B_{L_2})$ and $B_{L}=(B_{L_1}, B_{L_2})$ coincide with the coefficients $\tilde B_{L}$ and $B_{L}$ computed in section \ref{sec:Example2}. In addition, for each decay we could have in principle different spin analysers $\alpha$ and $\delta$ parameters, so
\begin{equation}
\begin{array}{ccc}
   \tilde B_{0}=1,  & \tilde B_{L_i=1}=\sqrt{\dfrac{3}{2}}\alpha_i, & \tilde B_{L_i=2}=\dfrac{(1-3\delta_i)}{\sqrt{2}}, \\ \ecart
    B_{0}=\sqrt{4\pi}, & B_{L_i=1}=\sqrt{2\pi}\alpha_i, & B_{L_i=2}=\sqrt{\dfrac{2\pi}{5}}(1-3\delta_i).
\end{array}
\end{equation}
Finally, replacing their values in Eq.\eqref{QTrhoProd}:
\begin{equation}
\begin{array}{c}
\dis A_{1\,M_1,\,0\,0}=\dfrac{\sqrt{6}}{\alpha_1}\int d\Omega_1 d\Omega_2\, \left[\dfrac{1}{\sigma}\dfrac{d \sigma}{d\Omega_1 d\Omega_2}\right]D^1_{M_1\,0}\lp \Omega_1\rp, \bigskip\\
A_{2\,M_1,\,0\,0}=\dfrac{5\sqrt{2}}{1-3\delta_1}\int d\Omega_1 d\Omega_2\, \left[\dfrac{1}{\sigma}\dfrac{d \sigma}{d\Omega_1 d\Omega_2}\right]D^2_{M_1\,0}\lp \Omega_1\rp, \bigskip \\ 
\dis A_{1\,M_1,\,1\,M_2}=\dfrac{6}{\alpha_1\alpha_2}\int d\Omega_1 d\Omega_2\, \left[\dfrac{1}{\sigma}\dfrac{d \sigma}{d\Omega_1 d\Omega_2}\right]D^1_{M_1\,0}\lp \Omega_1\rp\,D^1_{M_2\,0}\lp \Omega_2\rp,\bigskip  \\ 
\dis A_{1\,M_1,\,2\,M_2}=\dfrac{10\sqrt{3}}{\alpha_1(1-3\delta_2)}\int d\Omega_1 d\Omega_2\, \left[\dfrac{1}{\sigma}\dfrac{d \sigma}{d\Omega_1 d\Omega_2}\right]D^1_{M_1\,0}\lp \Omega_1\rp\,D^2_{M_2\,0}\lp \Omega_2\rp,\bigskip  \\ 
\dis A_{2\,M_1,\,2\,M_2}=\dfrac{50}{(1-3\delta_1)(1-3\delta_2)}\int d\Omega_1 d\Omega_2\, \left[\dfrac{1}{\sigma}\dfrac{d \sigma}{d\Omega_1 d\Omega_2}\right]D^2_{M_1\,0}\lp \Omega_1\rp\,D^2_{M_2\,0}\lp \Omega_2\rp
\end{array}
\end{equation}
and similarly for $A_{\,0\,0,\,L_2\,M_2}$ and $ A_{2\,M_1,\,1\,M_2}$ ($A_{0\,0,\,0\,0}=1$ due to normalisation).
We will always consider the final fermions massless in comparison to the mass of the initial bosons, therefore $\delta_i=0$, the value of $\alpha$ on the contrary will depend on the vector boson in hand. Namely, for the $Z$ boson $\alpha_Z\simeq-0.13$ while for the $W$ boson $\alpha_W\simeq-1$.

\subsection{$t\bar t W\to (bW^+) (\bar b W^-) (l\nu_l) \to (bl^+\nu_l) (\bar b l^-\bar \nu_l) (l \nu_l)$}\label{sec:Example5}
In this final example we cover the process for which 3 different particles reach a final 8-particle state. We aim to illustrate a constructive approach to measure the spin correlation tensor between 3 particles (among all the others spin polarisation and spin correlation matrices). Since each decay takes place independently, the production matrix for the whole process factorises as
\begin{equation}
    \Gamma^T(\theta_1,\varphi_1,\theta_2,\varphi_2,\theta_3,\varphi_3)=\Gamma_t^T(\theta_1,\varphi_1)\otimes \Gamma_{\bar t}^T(\theta_2,\varphi_2)\otimes \Gamma_{W}^T(\theta_3,\varphi_3).
\end{equation}
The first two production matrices coincide in form with the one of spin-$1/2$ particles, Eq.\eqref{eq:Gamma1/2}, while the latter one corresponds to the production matrix for a spin-1 particle, Eq.\eqref{eq:Gamma1}. The angular coordinates appearing in the argument of each matrix identify, for each particular decay, the momentum direction of the final charged lepton.

Analogously to the previous example, the components of the vectors $\tilde B_{L}=(\tilde B_{L_1},\tilde B_{L_2},\tilde B_{L_3})$ and $B_{L}=( B_{L_1}, B_{L_2},B_{L_3})$ are directly computed by replacing the corresponding values of $\alpha$ and $\delta$ in the expressions \eqref{BCoef1/2} and \eqref{BCoef1}. For instance, in the top and antitop decay we have $\alpha_t,\alpha_{\bar t}\simeq1$, whereas for the $W$ decay $\alpha_W\simeq-1$ and $\delta=0$. Using these specific values we get  
\begin{equation}
\begin{array}{ccc}
   \tilde B_{0}=1,  & \tilde B_{L_1=1}=\tilde B_{L_2=1}=1, & \tilde B_{L_3=1}=-\sqrt{\dfrac{3}{2}},\quad \tilde B_{L_3=2}=\dfrac{1}{\sqrt{2}}, \\ \ecart
    B_{0}=\sqrt{4\pi},  &  B_{L_1=1}=B_{L_2=1}=\sqrt{\dfrac{4\pi}{3}}, & B_{L_3=1}=-\sqrt{2\pi},\quad  B_{L_3=2}=\sqrt{\dfrac{2\pi}{5}}.
\end{array}
\end{equation}
The expressions for the coefficients entering in $\rho$ decomposition are:
\begin{equation*}
\begin{array}{c}
   \dis A_{L_1\,M_1,\,L_2\,M_2,\,L_3\,M_3}=\dfrac{(4\pi)^3}{B_{L_1}B_{L_2}B_{L_3}}\int d\Omega_1 d\Omega_2 d\Omega_3\left[\dfrac{1}{\sigma}\dfrac{d \sigma}{d\Omega_1 d\Omega_2 d\Omega_3}\right] \times \\ \ecart \dis
      \left[\lp\dfrac{(2L_1+1)(2L_2+1)(2L_3+1)}{4\pi}\rp^{1/2}D^{L_1}_{M_1\,0}\lp \Omega_1\rp D^{L_2}_{M_2\,0}\lp \Omega_2\rp D^{L_3}_{M_3\,0}\lp \Omega_3\rp\right].
\end{array}
\end{equation*}
In the latter equation we have not replaced the different values of $B_{L_i}$ and $L_i$ in order to give a more compact expression, but further simplifications take place when replacing them. The values of $A_{L_1\,M_1,\,L_2\,M_2,\,L_3\,M_3}$ associated with $L_i\neq0\quad \forall\,i$ conform the spin correlation tensor of the 3-particle system, while analogous interpretations for the spin correlation matrix and spin polarisation vectors as the ones given in the previous example apply for $A_{L_1\,M_1,\,L_2\,M_2,\,0\,0}$, $A_{L_1\,M_1,\,0\,0,\,L_3\,M_3}$, etc.

\section{Weyl-Wigner-Moyal formalism}\label{sec:WignerPQ}

In this section, we generalise the definition of Wigner Q and P symbols of an operator given a projective measure \cite{Li_2013} to account for operators acting over larger Hilbert spaces and when considering non-projective measures. Once this generalisation is set, we explain the relation between these concepts and the quantum tomography procedure we have developed, deepening in the relation between Particle Physics and Quantum Information first noticed in \cite{Ashby_Pickering_2023}

We start by introducing the whole formalism of the Wigner $P$ and $Q$ symbols. Let $\mathbf{J}$ be the vector spin operator fulfilling $\mathbf{J}\cdot \mathbf{J}=j(j+1)$, then the original definition of the Wigner $Q$ symbol associated with an operator $A\lp \mathbf{J}\rp$ acting on a particle with definite spin $j$ is that of the function
\begin{equation}
    \Phi^Q_A\lp\hat{n}\rp=\bra{\hat{n}}A\lp \mathbf{J}\rp\ket{\hat{n}}=\Tr{A\lp \mathbf{J}\rp \Pi_{\hat{n}}\lp \theta, \varphi\rp}.
\end{equation}
Here $\hat{n}=\hat{n}\lp \theta, \varphi\rp$ is a three dimensional unit vector characterised by the spherical coordinates $\lp \theta, \varphi\rp$, $\ket{\hat{n}}$ is the normalised state satisfying $\mathbf{J}\cdot\hat{n}\ket{\hat{n}}=j\ket{\hat{n}}$ and $\Pi_{\hat{n}}\lp \theta, \varphi\rp=\ket{\hat{n}}\bra{\hat{n}}$ is a projective measure over that state. Meanwhile, the corresponding Wigner $P$ symbol (in general not unique) for the operator $A\lp \mathbf{J}\rp$ is defined by 
\begin{equation}
    A\lp \mathbf{J}\rp=\frac{2j+1}{4\pi}\int d\hat{n}\, \Phi^P_A\lp\hat{n}\rp \ket{\hat{n}}\bra{\hat{n}}=\frac{2j+1}{4\pi}\int d\Omega_{\vec{n}}\, \Phi^P_A\lp\Omega_{\vec{n}}\rp \Pi_{\hat{n}}\lp\Omega_{\vec{n}}\rp,
\end{equation}
where $\Omega_{\vec{n}}$ is the solid angle defined by $\lp \theta, \varphi\rp$. An equivalent definition for the Wigner $P$ symbol of an operator $A\lp \mathbf{J}\rp$ is that of a function $\Phi^P_A\lp\hat{n}\rp$ satisfying that for any other operator $B\lp \mathbf{J}\rp$ the following property holds:
\begin{equation}
    \Tr{A\lp \mathbf{J}\rp\,B\lp \mathbf{J}\rp}=\frac{2j+1}{4\pi}\int d\hat{n}\, \Phi^P_A\lp\hat{n}\rp\Phi^Q_B\lp\hat{n}\rp=\frac{2j+1}{4\pi}\int d\Omega_{\vec{n}}\,\Phi^P_A\lp\Omega_{\vec{n}}\rp \Phi^Q_B\lp\Omega_{\vec{n}}\rp.
\end{equation}
These two notions are generalised when considering operators acting on larger Hilbert spaces, $A\lp\Omega,\bar{\kappa},\kappa\rp$, and when dealing with non-projective measures, $F\lp\Omega,\bar{\kappa},\kappa\rp$. The parameters $\Omega$, $\bar{\kappa}$ and $\kappa$ refer to the ones presented in previous sections and, unless it is needed, will be omitted in what follows. In particular, the generalised Wigner $Q$ symbol of an operator $A$ with respect to $F$ is
\begin{equation}
    \tilde{\Phi}^Q_A\equiv\Tr{A\, F}.
\end{equation}
The operator $F$ corresponds to a Positive Operator-Valued Measure, POVM
\cite{Ashby_Pickering_2023,nielsen_chuang_2010}. This is an element of a set of positive semi-definite Hermitian operators $\{F_l=\mathcal{K}_l^{\dagger}\mathcal{K}_l\}_l$, where the so-called Kraus operators $\mathcal{K}_l$ \cite{nielsen_chuang_2010} fulfil 
\begin{equation}
    \sum_l \mathcal{K}_l^{\dagger}\mathcal{K}_l=\sum_l F_l=\mathbb{1}.
\end{equation}
In our context and according to the definition given in \cite{Ashby_Pickering_2023}, the components of the measurement operator $\mathcal{K}$ leading to  $F=\mathcal{K}^{\dagger}\mathcal{K}$ coincide, up to a normalisation factor, with the helicity amplitudes ${\cal M}_{\lambda\,\bar{\lambda}}$ introduced in section \ref{sec:GenProdMat}. Hence,
\begin{equation}
    \mathcal{K}_{\lambda\,\bar{\lambda}}\propto{\cal M}_{\lambda\,\bar{\lambda}}\implies F_{\bar{\lambda}\,\bar{\lambda}'}\propto\sum_\lambda {\cal M}_{\lambda\,\bar{\lambda}}^* {\cal M}_{\lambda\,\bar{\lambda}'}.
\end{equation}
In consequence, comparing this result to the expression of $\Gamma^T$ and taking into account the normalisation factor, we obtain
\begin{equation}\label{GammaPOVM}
F_{\bar{\lambda}\,\bar{\lambda}'}=\Gamma^T_{\bar{\lambda}\,\bar{\lambda}'}\implies\tilde{\Phi}^Q_A=\Tr{A\, \Gamma^T}.
\end{equation}
Therefore, the POVM $F$ is exactly the transposed of the production matrix related to the scattering process, $\Gamma^T$. The interpretation of $\Gamma^T$ as a POVM may be tricky, as for it to hold one needs by definition a \textit{set} of positive semi-definite Hermitian operators $\{F_l\}_l$ such that $\sum_l F_l=\mathbb{1}$, while in our case we are dealing with a single one. However, one can always complete the set by considering only two POVM, $\{\Gamma^T,\mathbb{1}-\Gamma^T\}$.

Regarding the generalised Wigner $P$ symbol of an operator $A$, this one is defined as the function $\tilde{\Phi}^P_A$ (again not unique) such that for any other operator $B$ (or for a basis of operators) fulfils
\begin{equation}\label{GenPDef}
    \Tr{A\,B}=\frac{d}{8\pi^2 \bar{K} K}\int d\Omega\, \tilde{\Phi}^Q_B\, \tilde{\Phi}^P_A.
\end{equation}

Let us now check that we reproduce the quantum tomography of $\rho$ by only using the properties of these $Q$ and $P$ symbols, providing in addition explicit expressions for them in terms of Wigner D-matrices. 

For that purpose, denoting by $\lp\tilde{\Phi}^Q_{L\,M}\rp^{\lp \dagger \rp}$ and $\lp\tilde{\Phi}^P_{L\,M}\rp^{\lp \dagger \rp}$ the $Q$ and $P$ symbols of $\lp T^L_M\rp^{(\dagger)}$, it is clear from the results derived in section \ref{sec:DensMat} that
\begin{equation}
    \boxed{\tilde{\Phi}^Q_{L\,M}\lp\Omega,\bar{\kappa}, \kappa\rp=\Tr{T_M^L\, \Gamma^T\lp\Omega,\bar{\kappa}, \kappa\rp}=\sum_{M'}\tilde{B}_{L\,M'}\lp\bar{\kappa}, \kappa\rp^* D^L_{M\,M'}\lp \Omega\rp^*.}
\end{equation}
Moreover, we notice that when dealing with $A=T_M^L$ and $B=\lp T_{\hat{M}}^{\hat{L}}\rp^\dagger$, an equivalent formulation for the definition of the generalised $P$ symbol is deduced:
\begin{eqnarray}
    &\dis{d\,\delta_{L,\hat{L}}\, \delta_{M,\hat{M}}=\Tr{T_M^L\,\lp T_{\hat{M}}^{\hat{L}}\rp^\dagger}=\frac{d}{8\pi^2 \bar{K} K}\int d\Omega\, \tilde{\Phi}^Q_{L\,M}\, \lp\tilde{\Phi}^P_{\hat{L}\,\hat{M}}\rp^{\dagger}
    }\\ \ecart\nn
    &\dis{\iff \frac{1}{8\pi^2 \bar{K} K}\int d\Omega\, \tilde{\Phi}^Q_{L\,M}\, \lp\tilde{\Phi}^P_{\hat{L}\,\hat{M}}\rp^{\dagger}=\delta_{L\,\hat{L}}\, \delta_{M\,\hat{M}}.
    }
\end{eqnarray}
Thus, once the generalised $Q$ symbol for $T_M^L$ is known, a suitable family (labelled by $\hat{M}'$) of generalised $P$ symbols for $\lp T_M^L\rp^\dagger$  is
\begin{equation}
    \boxed{\lp\tilde{\Phi}^P_{\hat{L}\,\hat{M},\,\hat{M}'}\lp\Omega,\bar{\kappa}, \kappa\rp\rp^{\dagger}=\frac{4\pi}{B_{\hat{L}\,\hat{M}'}\lp\bar{\kappa}, \kappa\rp^*}\lp\frac{2\hat{L}+1}{4\pi}\rp^{1/2}D^{\hat{L}}_{\hat{M}\,\hat{M}'}\lp \Omega\rp,}
\end{equation}
which is well-defined provided that $B_{\hat{L}\,\hat{M}'}\neq0$. Indeed, we can verify this function satisfies the characterisation of the generalised $P$ symbol:
\begin{eqnarray}
    &\dis{\frac{1}{8\pi^2 \bar{K} K}\int d\Omega\, \tilde{\Phi}^Q_{L\,M}\, \lp\tilde{\Phi}^P_{\hat{L}\,\hat{M},\,\hat{M}'}\rp^{\dagger}=
    }\\ \ecart\nn
    &\dis{=\frac{1}{8\pi^2\bar{K} K}\int d\Omega\, \left[\sum_{M'}\tilde{B}_{L\,M'}^* D^L_{M\,M'}\lp \Omega\rp^*\right]\left[\frac{4\pi}{B_{\hat{L}\,\hat{M}'}^*}\lp\frac{2\hat{L}+1}{4\pi}\rp^{1/2}D^{\hat{L}}_{\hat{M}\,\hat{M}'}\lp \Omega\rp\right]
    }\\ \ecart\nn
    &\dis{=\sum_{M'}\left[\lp\frac{4\pi}{2\hat{L}+1}\rp^{1/2} \frac{\tilde{B}_{L\,M'}^*}{\bar{K} K}\dfrac{1}{B_{\hat{L}\,\hat{M}'}^*}\right]\left[\dfrac{2\hat{L}+1}{8\pi^2}\,\int d\Omega\, D^L_{M\,M'}\lp \Omega\rp^*\,D^{\hat{L}}_{\hat{M}\,\hat{M}'}\lp \Omega\rp\right]
    }\\ \ecart\nn
    &\dis{=\sum_{M'}\left[\frac{B_{\hat{L}\,\hat{M}'}^*}{B_{\hat{L}\,\hat{M}'}^*}\right]\,\delta_{L\,\hat{L}}\, \delta_{M\,\hat{M}}\,\delta_{M'\,\hat{M}'}=\delta_{L\,\hat{L}}\, \delta_{M\,\hat{M}}.
    }
\end{eqnarray}
Concerning the density matrix $\rho$, applying the definition of the generalised $P$ symbol \eqref{GenPDef} for $A=\rho$ and $B=\lp T_{\hat{M}}^{\hat{L}}\rp^\dagger$, we get on the one hand the coefficient we want to compute
\begin{eqnarray}\label{QPSymbol1}
    &\dis{\frac{d}{8\pi^2 \bar{K} K}\int d\Omega\, \tilde{\Phi}^Q_{\rho}\, \lp\tilde{\Phi}^P_{\hat{L}\,\hat{M},\,\hat{M}'}\rp^{\dagger}= \Tr{\rho\, \lp T_{\hat{M}}^{\hat{L}}\rp^\dagger}=A_{\hat{L}\,\hat{M}},
    }
\end{eqnarray}
where in the last step we have used expansion \eqref{rhoExp}. It is left to relate the left-hand side of the previous equation to the angular distribution of the final particles. For instance, $\tilde{\Phi}^Q_{\rho}$ is directly related to the differential cross section by Eq.(\ref{MasterEq})
\begin{equation}
    \frac{1}{\sigma}\frac{d \sigma}{d\Omega\, d \bar{\kappa} \,d \kappa}=\frac{d}{8\pi^2 \bar{K} K}\Tr{\rho\, \Gamma^T}=\frac{d}{8\pi^2\bar{K} K}\tilde{\Phi}^Q_{\rho}\implies\tilde{\Phi}^Q_{\rho}=\frac{8\pi^2 \bar{K} K}{d} \frac{1}{\sigma}\frac{d \sigma}{d\Omega\, d \bar{\kappa} \,d \kappa}.
\end{equation}
Hence, using this characterisation of $\tilde{\Phi}^Q_{\rho}$ as well as the explicit form of $\lp\tilde{\Phi}^P_{\hat{L}\,\hat{M},\,\hat{M}'}\rp^{\dagger}$, we get on the other hand 
\begin{eqnarray}\label{QPSymbol2}
    &\dis{\frac{d}{8\pi^2 \bar{K}K}\int d\Omega\,\tilde{\Phi}^Q_{\rho}\, \lp\tilde{\Phi}^P_{\hat{L}\,\hat{M},\,\hat{M}'}\rp^{\dagger}=
    }\\ \ecart\nn
    &\dis{=\int d\Omega\, \left[\frac{1}{\sigma}\frac{d \sigma}{d\Omega\, d \bar{\kappa} \,d \kappa}\right]\, \lp\tilde{\Phi}^P_{\hat{L}\,\hat{M},\,\hat{M}'}\rp^{\dagger}=\frac{4\pi}{B_{\hat{L}\,\hat{M}'}^*}\int d\Omega\, \left[\frac{1}{\sigma}\frac{d \sigma}{d\Omega\, d \bar{\kappa} \,d \kappa}\right]\, \lp\frac{2\hat{L}+1}{4\pi}\rp^{1/2}D^{\hat{L}}_{\hat{M}\,\hat{M}'}\lp \Omega\rp.
    }
\end{eqnarray}
Finally, equating and simplifying both Eq.\eqref{QPSymbol1} and Eq.\eqref{QPSymbol2}, 
\begin{equation}
    \boxed{\int d\Omega\, \left[\frac{1}{\sigma}\frac{d \sigma}{d\Omega\, d \bar{\kappa} \,d \kappa}\right]\, \lp\frac{2\hat{L}+1}{4\pi}\rp^{1/2}D^{\hat{L}}_{\hat{M}\,\hat{M}'}\lp \Omega\rp=\frac{B_{\hat{L}\,\hat{M}'}^*}{4\pi}\,A_{\hat{L}\,\hat{M}}\lp\bar{\kappa}\rp.}
\end{equation}
This expression reproduces exactly the same result as the one derived in Eq.\eqref{QTrho}. For the choice $\hat{M}'=0$ when possible, the generalised $P$ symbol reduces to
\begin{equation}
    \lp\tilde{\Phi}^P_{\hat{L}\,\hat{M}}\rp^{\dagger}\equiv\lp\tilde{\Phi}^P_{\hat{L}\,\hat{M},\,0}\rp^{\dagger}=\frac{4\pi}{B_{\hat{L}\,0}}\lp\frac{2L+1}{4\pi}\rp^{1/2}D^{\hat{L}}_{\hat{M}\,0}\lp \Omega\rp=\frac{4\pi}{B_{\hat{L}}}Y_{\hat{L}}^{\hat{M}}\lp \theta,\varphi\rp^*.
\end{equation}

\section{Summary and conclusions} \label{sec:Conclusions}
In this paper we have provided a complete formalism for the quantum tomography of the helicity quantum initial state $\rho$ in a general scattering process. Following the method here developed, the coefficients in the parameterisation of the density matrix $\rho$ in terms of irreducible tensor operators $\{T^L_M\}$ are determined by averaging over the angular distribution data of the final state, see Eq.\eqref{QTrho}. As an intermediate step to accomplish the main goal, we have defined a generalisation of the production matrix of a scattering $\Gamma$, deriving its expansion with respect to the irreducible tensors (a formulation that makes explicit its angular dependence) as well as the exact form of its elements, which generalise the well-known reduced helicity amplitudes for a variety of processes (labelled according to the number of initial and final particles).

The crucial feature we exploited in order to achieve the tomography of the helicity initial state is the transformation property of the irreducible tensor basis under rotations of the system. Actually, this property along with the existent relation between $\rho$ and $\Gamma$ leads to a novel expansion of the normalised differential cross section in terms of the Wigner D-matrices, which eventually gives rise to a
constructive and experimentally practical algorithm to extract the relevant helicity information of any process in hand. Special attention has been given in this work to the case in which $\Gamma$ factorises, i.e. when the whole process can be decomposed in subgroups of independent processes. In particular, when all the involved scatterings are in fact decays, the basis coefficients in the $\rho$ expansion have a quite clear physical meaning and we recover and extend the studies previously done in the literature, which we illustrate with some simple examples.

Finally, we have also presented a re-derivation of all our previous results from a quantum-information perspective. Namely,  we based on the Weyl-Wigner-Moyal formalism to define and compute the generalised Wigner $P$ and $Q$ symbols for the irreducible tensor operators. Thus, compact analytical expression are deduced and a rather simple relation between Particle Physics and Quantum Information is manifested.

\section*{Acknowledgements}
The author is grateful to J. A. Aguilar-Saavedra, J.A. Casas and J. M. Moreno for very fruitful discussions. The author acknowledges the support of the Spanish Agencia Estatal de Investigacion through the grants ``IFT Centro de Excelencia Severo Ochoa CEX2020-001007-S" and PID2019-110058GB-C22 funded by MCIN/AEI/10.13039/501100011033 and by ERDF. The work of the author is supported through the FPI grant PRE2020-095867 funded by MCIN/AEI/10.13039/501100011033.

\newpage
\appendix

\section{Explicit computation of the generalised production matrix} \label{sec:CasesProdMat}
In this Appendix we compute in detail and for the cases mentioned in section \ref{sec:GenProdMat}, the expression of $a_{\sigma,\sigma'}$, $\tilde{B}_{L\,\sigma_T^{-}}$ and $B_{L\,M'}$. For the computation we will consider on-shell particles for the initial states as this feature is almost irrelevant for the reasoning. However, comments related to the off-shell case will be done when needed at the end of each subsection, as they may change the $B_{L\,M'}$ coefficient. Regarding the particles of the final state, they are always on-shell as they are the physical particles we are doing measurements on for reconstructing the density matrix $\rho$. Furthermore, we will make explicit the $\lp\bar{\mathbb{1}},\,\mathbb{1}\rp$ and $\lp\bar{\kappa},\kappa\rp$ dependence only when needed.

\subsection{$\bar{1}\to 2$}\label{sec:CasesProdMat12}
We recall that 
\begin{equation}
    \Gamma^T_{\bar{\lambda}\, \bar{\lambda}'}\propto \bra{\bar{\mathbb{1}}\,\bar{\kappa}\,\bar{\lambda}}\left[\sum_\lambda\lp S^\dagger\ket{\mathbb{1}\, \kappa\, \lambda} \bra{\mathbb{1}\, \kappa\, \lambda}S \rp \right]\ket{\bar{\mathbb{1}}\,\bar{\kappa}\,\bar{\lambda}'},
\end{equation}
with
\begin{eqnarray}
    \ket{\mathbb{1}\, \kappa\, \lambda}&\propto&\dis{\sum_J \sum_{M\,\Lambda}(2J+1)^{1/2}D^J_{M\,\Lambda}\lp\mathbb{1}\rp\ket{J\,M\,\Lambda\,\kappa\,\lambda}
    }\\ \ecart \nn
    &=&\dis{\sum_J \sum_{M\,\Lambda}(2J+1)^{1/2}\delta_{M\,\Lambda}\ket{J\,M\,\Lambda\,\kappa\,\lambda}=\sum_J \sum_{\Lambda}(2J+1)^{1/2}\ket{J\,\Lambda\,\kappa\,\lambda}.
    }
\end{eqnarray}

Due to the reduced number of particles in both the initial and final states, we have $\bar{\kappa}=\kappa=\varnothing$. Furthermore, $\bar{\Lambda}=\bar{\lambda}$ and $\Lambda= \lambda_1-\lambda_2\equiv\zeta_\lambda^{-}$ are well-defined quantum numbers, so we do not sum in them. In addition, the initial total angular momentum is also well-defined: $\bar{J}=\bar{s}$, where $\bar{s}$ is the spin of the initial particle. Taking everything into account:

\begin{equation}\left.\begin{array}{l}
    \ket{\bar{\mathbb{1}}\, \bar{\kappa}\, \bar{\lambda}}=\ket{\bar{\mathbb{1}}\,\bar{\lambda}}\propto\lp2\bar{s}+1\rp^{1/2}\ket{\bar{s}\,\bar{\lambda}}\propto\ket{\bar{s}\,\bar{\lambda}}\\ \ecart
    \ket{\mathbb{1}\, \kappa\, \lambda}=\ket{\mathbb{1}\,\lambda}\propto\dis\sum_{J}\lp 2J+1\rp^{1/2}\ket{J\,\zeta_\lambda^{-}\,\lambda}
    \end{array}\right\}\implies
\end{equation}
\begin{eqnarray}\nn
    \implies\Gamma^T_{\bar{\lambda}\, \bar{\lambda}'}&\propto&\dis{\bra{\bar{s}\,\bar{\lambda}}\sum_{\lambda_1\,\lambda_2} S^\dagger \lp\sum_{J}\lp 2J+1\rp^{1/2}\ket{J\,\zeta_\lambda^{-}\,\lambda}\rp\lp\sum_{J'}\lp 2J'+1\rp^{1/2}\bra{J'\,\zeta_\lambda^{-}\,\lambda}\rp S \ket{\bar{s}\,\bar{\lambda}'}
    }\\ \ecart \nn
    &=&\sum_{\lambda_1\,\lambda_2}\sum_{J\,J'}\lp 2J+1\rp^{1/2}\lp 2J'+1\rp^{1/2}\bra{\bar{s}\,\bar{\lambda}}S^\dagger\ket{J\,\zeta_\lambda^{-}\,\lambda}\bra{J'\,\zeta_\lambda^{-}\,\lambda}S\ket{\bar{s}\,\bar{\lambda}'}.
\end{eqnarray}
As the scattering $S-$matrix conserves both $J^2$ and $J_z$ (this last one coincides with $\Lambda$ because $D^J_{M\,\Lambda}\lp\mathbb{1}\rp=\delta_{M\,\Lambda}$), we have
\begin{equation}\left.\begin{array}{l}
    \bra{\bar{s}\,\bar{\lambda}}S^\dagger\ket{J\,\zeta_\lambda^{-}\,\lambda}=\bra{\bar{\lambda}}{S^{\bar{s}\ \dagger}_{\zeta_\lambda^{-}}}\ket{\lambda}\,\delta_{\bar{s}\,J}\,\delta_{\bar{\lambda}\,\zeta_\lambda^{-}}\\ \ecart
    \bra{J'\,\zeta_\lambda^{-}\,\lambda}S\ket{\bar{s}\,\bar{\lambda}'}=\bra{\lambda}{S^{\bar{s}}_{\zeta_\lambda^{-}}}\ket{\bar{\lambda}'}\,\delta_{\bar{s}\,J'}\,\delta_{\bar{\lambda}'\,\zeta_\lambda^{-}}
    \end{array}\right\}\implies
\end{equation}
\begin{eqnarray}\nn
    \implies\Gamma^T_{\bar{\lambda}\, \bar{\lambda}'}&\propto&\dis{ \sum_{\lambda_1\,\lambda_2}\bra{\bar{\lambda}}{S^{\bar{s}\ \dagger}_{\zeta_\lambda^{-}}}\ket{\lambda}\bra{\lambda}{S^{\bar{s}}_{\zeta_\lambda^{-}}}\ket{\bar{\lambda}'}\delta_{\bar{\lambda}\,\zeta_\lambda^{-}}\,\delta_{\bar{\lambda}'\,\zeta_\lambda^{-}}
    }\\ \ecart \nn
    &=&\sum_{\lambda_1\,\lambda_2}\left|\bra{\lambda}{S^{\bar{s}}_{\zeta_\lambda^{-}}}\ket{\bar{\lambda}}\right|^2 \delta_{\bar{\lambda}\,\bar{\lambda}'}\,\delta_{\bar{\lambda}\,\zeta_\lambda^{-}}.
\end{eqnarray}
Rewriting the sum in $\lp \lambda_1, \lambda_2\rp$ as one in  $\lp \lambda_1,\zeta_\lambda^{-}\rp$ and applying the Kronecker delta
\begin{eqnarray}
    &\dis{\Gamma^T_{\bar{\lambda}\, \bar{\lambda}'}\propto \sum_{\lambda_1}\sum_{\zeta_\lambda^{-}}\left|\bra{\lambda}{S^{\bar{s}}_{\zeta_\lambda^{-}}}\ket{\bar{\lambda}}\right|^2 \delta_{\bar{\lambda}\,\bar{\lambda}'}\,\delta_{\bar{\lambda}\,\zeta_\lambda^{-}}\propto
    \sum_{\zeta_\lambda^{-}}a_{\zeta_\lambda^{-}}\delta_{\bar{\lambda}\,\bar{\lambda}'}\,\delta_{\bar{\lambda}\,\zeta_\lambda^{-}}, 
    } \\ \ecart \nn
    &\dis{a_{\zeta_\lambda^{-}}=\dfrac{2\bar{s}+1}{4\pi} \sum_{\lambda_1}\left|\bra{\lambda_1\,\lambda_1-\zeta_\lambda^{-}}{S^{\bar{s}}_{\zeta_\lambda^{-}}}\ket{\zeta_\lambda^{-}}\right|^2,
    }
\end{eqnarray}
where we have introduced the well-known reduced helicity amplitudes \cite{haber1994spin,Boudjema_2009,Rahaman_2022}, that we are denoting here as $a_{\zeta_\lambda^{-}}$. Moreover, since the $\lp\bar{\lambda}, \bar{\lambda}'\rp$ element of $\Gamma^T$ is proportional to $\delta_{\bar{\lambda}\,\bar{\lambda}'}$, it is clear that $\Gamma^T$ is diagonal and its expression in terms of the $e_{\sigma\,\sigma'}$ matrices is
\begin{equation}
    \Gamma^T\propto \sum_{\sigma\,\sigma'}a_{\sigma\,\sigma'}\,e_{\sigma\,\sigma'},\hbox{ with } a_{\sigma\,\sigma'}=a_{\sigma}\,\delta_{\sigma\,\sigma'}=\dfrac{2\bar{s}+1}{4\pi} \sum_{\lambda_1}\left|\bra{\lambda_1\,\lambda_1-\sigma}{S^{\bar{s}}_{\sigma}}\ket{\sigma}\right|^2 \delta_{\sigma\,\sigma'}.
\end{equation}
Taking into account the normalisation factor $a_{+}=\displaystyle\sum_{\sigma}a_{\sigma\,\sigma}=\sum_{\sigma}a_{\sigma}$:
\begin{equation}
\Gamma^T=\dis{\dfrac{1}{a_{+}}\sum_{\sigma}a_{\sigma}\,e_{\sigma\,\sigma}} ,\hbox{ with } a_{\sigma}=\dfrac{2\bar{s}+1}{4\pi} \sum_{\lambda_1}\left|\bra{\lambda_1\,\lambda_1-\sigma}{S^{\bar{s}}_{\sigma}}\ket{\sigma}\right|^2.
\end{equation}

Finally, let us obtain $\tilde{B}_{L\,\sigma_T^{-}}$ and $B_{L\,M'}$ knowing $a_{\sigma\,\sigma'}$. Due to $m=1$, we have  $d^{(v)}=1$ and therefore $\sigma_T^{-}=\sigma-\sigma'=\sigma^-$ and $C\lp L,\,\bar{s},\,\sigma,\,\sigma'\rp=\lp 2L +1\rp ^{1/2} C^{\bar{s}\, \sigma}_{\bar{s}\, \sigma'\, L\, \sigma^{-}}$. Hence, substituting in the expression for $\tilde{B}_{L\,\sigma_T^{-}}=\tilde{B}_{L\,\sigma^{-}}$ and noticing that $\delta_{\sigma\,\sigma'}=\delta_{\sigma\,\sigma'}\,\delta_{\sigma^{-}\,0}$,
\begin{equation}
    \tilde{B}_{L\,\sigma^{-}}=\frac{1}{a_{+}}\sum_{\substack{\sigma\, \sigma'\\ \sigma-\sigma'=\sigma^{-}}}a_{\sigma\, \sigma}\,\delta_{\sigma\,\sigma'}\, \lp 2L +1\rp ^{1/2} C^{\bar{s}\, \sigma}_{\bar{s}\, \sigma'\, L\, \sigma^{-}}=\frac{\lp 2L +1\rp ^{1/2}}{a_{+}}\sum_{\sigma}a_{\sigma}\,C^{\bar{s}\, \sigma}_{\bar{s}\, \sigma\, L\, 0}\,\delta_{\sigma^{-}\,0}.
\end{equation}
In consequence, the only possible non-zero $\tilde{B}_{L\,\sigma^{-}}$ is $\tilde{B}_{L\,0}$. For $B_{L\,M'}$, one should distinguish between the on-shell and off-shell cases for the particle in the initial state, as it leads to $\bar{\kappa}=\varnothing$ and $\bar{\kappa}=\{\bar{m}\}$ respectively. Namely, for the on-shell scenario
\begin{equation}
    B_{L\,M'}=\lp\frac{4\pi}{2L+1}\rp^{1/2} \tilde{B}_{L\,M'}=\lp\frac{4\pi}{2L+1}\rp^{1/2} \tilde{B}_{L\,0}\,\delta_{M'\,0}=\frac{\sqrt{4\pi}}{a_{+}}\sum_{\sigma} a_{\sigma}\,C^{\bar{s}\, \sigma}_{\bar{s}\, \sigma\, L\, 0}\,\delta_{M'\,0}=B_{L}\,\delta_{M'\,0}.
\end{equation}
Meanwhile, for the off-shell one 
\begin{equation}
    B_{L\,M'}=\lp\frac{4\pi}{2L+1}\rp^{1/2} \frac{\tilde{B}_{L\,M'}\lp\bar{m}\rp}{\bar{K}}=\frac{\sqrt{4\pi}}{a_{+}\lp\bar{m}\rp\bar{K}}\sum_{\sigma} a_{\sigma}\lp\bar{m}\rp\,C^{\bar{s}\, \sigma}_{\bar{s}\, \sigma\, L\, 0}\,\delta_{M'\,0}, \quad \hbox{ with }\bar{K}=\int d\,\bar{m}.
\end{equation}
Furthermore, due to the normalisation condition over $\Gamma^T$, all the magnitudes appearing in $B_{L\,M'}$ should come in a dimensionless combination. In the special case of massless final particles, $\bar{m}$ becomes the only dimensional 
free magnitude and therefore can not appear in $B_{L\,M'}$. Actually for the massive case, the $B_{L\,M'}$ dependence on the masses of the particles comes as the combination $m_i/\bar{m}$, with $m_i$ the masses of the on-shell final particles.  This is indeed what has been obtained in refs. \cite{Boudjema_2009,Rahaman_2022,PhysRevD.107.016012}.

\subsection{$\bar{1}\to n$}\label{sec:CasesProdMat1n}

As the initial state is the same than in the previous case, we still have $\bar{\kappa}=\varnothing$, $\bar{J}=\bar{s}$ and $\bar{\Lambda}=\bar{\lambda}$, leading to
\begin{equation}
    \ket{\bar{\mathbb{1}}\, \bar{\kappa}\, \bar{\lambda}}\propto\ket{\bar{s}\,\bar{\lambda}}.
\end{equation}
On the other hand, for the final state it will no longer be true that $\Lambda$ is well-defined nor $\kappa=\varnothing$. Then,
\begin{equation}
    \ket{\mathbb{1}\, \kappa\, \lambda}\propto\sum_{J}\sum_{\Lambda}\lp 2J+1\rp^{1/2}\ket{J\,\Lambda\,\kappa\,\lambda}.
\end{equation}
Following the same reasoning done before
\begin{equation}
    \Gamma^T_{\bar{\lambda}\, \bar{\lambda}'}\propto\sum_{\lambda}\sum_{J\,J'}\sum_{\Lambda\,\Lambda'}\lp 2J+1\rp^{1/2}\,\lp 2J'+1\rp^{1/2}\bra{\bar{s}\,\bar{\lambda}}S^\dagger\ket{J\,\Lambda\,\kappa\,\lambda}\bra{J'\,\Lambda'\,\kappa\,\lambda}S\ket{\bar{s}\,\bar{\lambda}'}.
\end{equation}
Applying conservation of both $J^2$ and $J_z$ (this last one coincides with $\Lambda$ because $D^J_{M\,\Lambda}\lp\mathbb{1}\rp=\delta_{M\,\Lambda}$):
\begin{equation}\left.\begin{array}{l}
    \lp 2J+1\rp^{1/2}\bra{\bar{s}\,\bar{\lambda}}S^\dagger\ket{J\,\Lambda\,\kappa\,\lambda}=\lp 2\bar{s}+1\rp^{1/2}\bra{\bar{\lambda}}{S^{\bar{s}\ \dagger}_{\Lambda}}\ket{\kappa\,\lambda}\,\delta_{\bar{s}\,J}\,\delta_{\bar{\lambda}\,\Lambda}\\ \ecart
     \lp 2J'+1\rp^{1/2}\bra{J'\,\Lambda'\,\kappa\,\lambda}S\ket{\bar{s}\,\bar{\lambda}'}=\lp 2\bar{s}+1\rp^{1/2}\bra{\kappa\,\lambda}{S^{\bar{s}}_{\Lambda'}}\ket{\bar{\lambda}'}\,\delta_{\bar{s}\,J'}\,\delta_{\bar{\lambda}'\,\Lambda'}
    \end{array}\right\}\implies
\end{equation}
\begin{eqnarray}\nn
    &\implies\Gamma^T_{\bar{\lambda}\, \bar{\lambda}'}\propto\dis{\sum_{\Lambda\,\Lambda'}\sum_{\lambda}\bra{\Lambda}{S^{\bar{s}\ \dagger}_{\Lambda}}\ket{\kappa\,\lambda}\bra{\kappa\,\lambda}{S^{\bar{s}}_{\Lambda'}}\ket{\Lambda'}\,\delta_{\bar{\lambda}\,\Lambda}\,\delta_{\bar{\lambda}'\,\Lambda'}\propto\sum_{\Lambda\,\Lambda'}a_{\Lambda\,\Lambda'}\,\delta_{\bar{\lambda}\,\Lambda}\,\delta_{\bar{\lambda}'\,\Lambda'},
    }\\ \ecart \nn
    &\dis{a_{\Lambda\,\Lambda'}=\dfrac{2\bar{s}+1}{4\pi} \sum_{\lambda}\bra{\Lambda}{S^{\bar{s}\ \dagger}_{\Lambda}}\ket{\kappa\,\lambda}\bra{\kappa\,\lambda}{S^{\bar{s}}_{\Lambda'}}\ket{\Lambda'}.
    }
\end{eqnarray}
The expression of $\Gamma^T$ in the $e_{\sigma\,\sigma}$ basis is straightforward:
\begin{eqnarray}
    &\dis{\Gamma^T=\dfrac{1}{a_{+}}\sum_{\sigma\,\sigma'}a_{\sigma\,\sigma'}\,e_{\sigma\,\sigma'},
    }\\ \ecart \nn
    &\dis{\hbox{ with } a_{\sigma\,\sigma'}=\dfrac{2\bar{s}+1}{4\pi} \sum_{\lambda}\bra{\sigma}{S^{\bar{s}\ \dagger}_{\sigma}}\ket{\kappa\,\lambda}\bra{\kappa\,\lambda}{S^{\bar{s}}_{\sigma'}}\ket{\sigma'} \hbox{ and } a_{+}=\sum_{\sigma}a_{\sigma\,\sigma}.
    }
\end{eqnarray}
Like in the $\bar{1}\to2$ case, $\sigma_T^{-}=\sigma-\sigma'=\sigma^-$ and $C\lp L,\,\bar{s},\,\sigma,\,\sigma'\rp=\lp 2L +1\rp ^{1/2} C^{\bar{s}\, \sigma}_{\bar{s}\, \sigma'\, L\, \sigma^{-}}$, simplifying the explicit form of $\tilde{B}_{L\,\sigma_T^{-}}$ and $B_{L\,M'}$. Namely, taking into account that $\kappa\neq\varnothing$,
\begin{eqnarray}
    &\dis{\tilde{B}_{L\,\sigma^{-}}=\frac{\lp 2L +1\rp ^{1/2}}{a_{+}}\sum_{\substack{\sigma\, \sigma'\\ \sigma-\sigma'=\sigma^{-}}}a_{\sigma\, \sigma'}\, C^{\bar{s}\, \sigma}_{\bar{s}\, \sigma'\, L\, \sigma^{-}}=\frac{\lp 2L +1\rp ^{1/2}}{a_{+}}\sum_{\sigma}a_{\sigma\,\lp\sigma-\sigma^{-}\rp}\,C^{\bar{s}\, \sigma}_{\bar{s}\, \lp\sigma-\sigma^{-}\rp\, L\, \sigma^{-}},
    }\\ \ecart \nn
    &\dis{B_{L\,M'}=\lp\frac{4\pi}{2L+1}\rp^{1/2}\frac{\tilde{B}_{L\,M'}}{K} =\frac{\sqrt{4\pi}}{ a_+ K}\sum_{\sigma}a_{\sigma\, \lp\sigma-M'\rp}\,C^{\bar{s}\, \sigma}_{\bar{s}\, \lp\sigma-M'\rp\, L\, M'}.
    }
\end{eqnarray}
Regarding the $M'=0$ choice,
\begin{equation}
    B_{L}=\frac{\sqrt{4\pi}}{a_+ K} \sum_{\sigma}a_{\sigma\, \sigma}\,C^{\bar{s}\, \sigma}_{\bar{s}\, \sigma\, L\, 0},\hbox{ with } a_{\sigma\,\sigma}=\dfrac{2\bar{s}+1}{4\pi} \sum_{\lambda}\left|\bra{\kappa\,\lambda}{S^{\bar{s}}_{\sigma}}\ket{\sigma}\right|^2.
\end{equation}
In contrast, when the initial particle is off-shell, a dependence on the initial mass appears. Therefore, $\bar{\kappa}\neq\varnothing$ and
\begin{eqnarray}
    &\dis{B_{L\,M'}=\lp\frac{4\pi}{2L+1}\rp^{1/2}\frac{\tilde{B}_{L\,M'}}{\bar{K} K} =\frac{\sqrt{4\pi}}{ a_+ \bar{K} K}\sum_{\sigma}a_{\sigma\, \lp\sigma-M'\rp}\,C^{\bar{s}\, \sigma}_{\bar{s}\, \lp\sigma-M'\rp\, L\, M'},
    }\\ \ecart \nn
    &\dis{B_{L}=\frac{\sqrt{4\pi}}{a_+ \bar{K} K} \sum_{\sigma}a_{\sigma\, \sigma}\,C^{\bar{s}\, \sigma}_{\bar{s}\, \sigma\, L\, 0}.
    }
\end{eqnarray}

\subsection{$\bar{2}\to 2$}\label{sec:CasesProdMat22}
For this case, neither the initial nor final states have well-defined $J^2$. Nevertheless, it is still true that $J_z$ is a well-defined quantum number. Moreover, $\bar{\kappa}=\kappa=\varnothing$. Denoting $\bar{\zeta}^{-}_{\lambda}\equiv\bar{\lambda}_1-\bar{\lambda}_2$ and $\zeta^{-}_{\lambda}\equiv\lambda_1-\lambda_2$ and following the standard steps:
\begin{equation}\left.\begin{array}{l}
    \ket{\bar{\mathbb{1}}\, \bar{\kappa}\, \bar{\lambda}}=\ket{\bar{\mathbb{1}}\,\bar{\lambda}}\propto\dis\sum_{\bar{J}}\lp 2\bar{J}+1\rp^{1/2}\ket{\bar{J}\,\bar{\zeta}^{-}_{\lambda}\,\bar{\lambda}}\\ \ecart
    \ket{\mathbb{1}\, \kappa\, \lambda}=\ket{\mathbb{1}\,\lambda}\propto\dis\sum_{J}\lp 2J+1\rp^{1/2}\ket{J\,\zeta_\lambda^{-}\,\lambda}
    \end{array}\right\}\implies
\end{equation}
\begin{equation}
    \implies\Gamma^T_{\bar{\lambda}\, \bar{\lambda}'}\propto\sum_{\lambda_1\,\lambda_2}\sum_{J\,J'}\lp 2J+1\rp\,\lp 2J'+1\rp\bra{\bar{\lambda}}{S^{J\ \dagger}_{\zeta_\lambda^{-}}}\ket{\lambda}\bra{\lambda}{S^{J'}_{\zeta_\lambda^{-}}}\ket{\bar{\lambda}'}\, \delta_{\bar{\zeta}^{-}_{\lambda}\,\zeta_\lambda^{-}}\delta_{\bar{\zeta}^{-\,'}_{\lambda}\,\zeta_\lambda^{-}}.
\end{equation}
Thus,
\begin{equation}
    \Gamma^T=\dfrac{1}{a_{+}}\sum_{\sigma\,\sigma'}a_{\sigma\,\sigma'}\,e_{\sigma\,\sigma'},\hbox{ with } a_{+}=\sum_{\sigma}a_{\sigma\,\sigma}
\end{equation}
and the factor $a_{\sigma\,\sigma'}$ is given by
\begin{eqnarray}\nn
    a_{\sigma\,\sigma'}&=&\dis{\sum_{J\,J'}\lp\dfrac{2J+1}{4\pi}\rp\lp\dfrac{2J'+1}{4\pi}\rp\sum_{\lambda_1\,\lambda_2}\bra{\sigma_1\,\sigma_2}{S^{J\ \dagger}_{\zeta_\lambda^{-}}}\ket{\lambda_1\,\lambda_2}\bra{\lambda_1\,\lambda_2}{S^{J'}_{\zeta_\lambda^{-}}}\ket{\sigma_1',\sigma_2'}\, \delta_{\zeta^{-}_{\sigma}\,\zeta_\lambda^{-}}\delta_{\zeta^{-}_{\sigma}\,\zeta^{-\,'}_{\sigma}}
    }\\ \ecart \nn
    &=&\sum_{J\,J'}\lp\dfrac{2J+1}{4\pi}\rp\lp\dfrac{2J'+1}{4\pi}\rp\sum_{\lambda_1}\bra{\sigma_1\,\sigma_2}{S^{J\ \dagger}_{\zeta_\sigma^{-}}}\ket{\lambda_1\,\lp\lambda_1-\zeta_\sigma^{-}\rp}\bra{\lambda_1\,\lp\lambda_1-\zeta_\sigma^{-}\rp}{S^{J'}_{\zeta_\sigma^{-}}}\ket{\sigma_1',\sigma_2'}\,\delta_{\zeta^{-}_{\sigma}\,\zeta^{-\,'}_{\sigma}}.
\end{eqnarray}

Finally, let us obtain $\tilde{B}_{L\,\sigma_T^{-}}$ and $B_{L\,M'}$ knowing the structure of $a_{\sigma\,\sigma'}$. As stated in the main text, in $\tilde{B}_{L\,\sigma_T^{-}}$ we need to sum over $\sigma=\lp\sigma_1,\sigma_2\rp$ and $\sigma'=\lp\sigma_1',\sigma_2'\rp$ constrained to $\lp\sigma-\sigma'\rp\cdot d^{(v)}=\sigma_T-\sigma_T'=\sigma_T^{-}$. Actually, this condition can be rewritten as a Kronecker delta in a free sum over $\lp\sigma,\sigma'\rp$. In addition, using the restriction $\zeta^{-}_{\sigma}=\zeta^{-\,'}_{\sigma}$ coming from the $a_{\sigma\,\sigma'}$ expression, it is possible to rearrange the sums in the $\tilde{B}_{L\,\sigma_T^{-}}$ as
\begin{equation}
    \sum_{\sigma\,\sigma'}\delta_{\sigma_T'\,\lp\sigma_T-\sigma_T^{-}\rp}\delta_{\zeta^{-}_{\sigma}\,\zeta^{-\,'}_{\sigma}}\longrightarrow\sum_{\sigma_T\,\sigma_T'}\sum_{\zeta^{-}_{\sigma}\,\zeta^{-\,'}_{\sigma}}\delta_{\sigma_T'\,\lp\sigma_T-\sigma_T^{-}\rp}\delta_{\zeta^{-}_{\sigma}\,\zeta^{-\,'}_{\sigma}}\longrightarrow\sum_{\sigma_T\,\zeta^{-}_{\sigma}}\longrightarrow\sum_{\sigma}.
\end{equation}
Hence, instead of summing over both $\lp\sigma,\sigma'\rp$ it is only needed to sum over $\sigma=\lp\sigma_1,\sigma_2\rp$ once we have done the following replacements in $\sigma'=\lp\sigma_1',\sigma_2'\rp$:
\begin{eqnarray}
    &\dis{\sigma_1'=\frac{\sigma_T'+\zeta^{-\,'}_{\sigma}}{d_2+1}=\frac{\sigma_T+\zeta^{-}_{\sigma}}{d_2+1}-\frac{\sigma_T^{-}}{d_2+1}=\sigma_1-\frac{\sigma_T^{-}}{d_2+1},
    }\\ \ecart \nn
    &\dis{\sigma_2'=\frac{\sigma_T'- d_2\,\zeta^{-\,'}_{\sigma}}{d_2+1}=\frac{\sigma_T- d_2\,\zeta^{-}_{\sigma}}{d_2+1}-\frac{\sigma_T^{-}}{d_2+1}=\sigma_2-\frac{\sigma_T^{-}}{d_2+1}.
    }
\end{eqnarray}
In consequence, one has
\begin{eqnarray}
    &\dis{\tilde{B}_{L\,\sigma^{-}_T}=\frac{\lp 2L +1\rp ^{1/2}}{a_{+}}\sum_{\sigma_1\,\sigma_2}\left.a_{\sigma\, \sigma'}\right|_{\sigma^{-}_T}\, C^{s_T\, \sigma_T}_{s_T\, \lp\sigma_T-\sigma_T^{-}\rp\, L\, \sigma_T^{-}},
    }\\ \ecart \nn
    &\dis{B_{L\,M'}=\lp\frac{4\pi}{2L+1}\rp^{1/2} \tilde{B}_{L\,M'}=\frac{\sqrt{4\pi}}{a_+}\sum_{\sigma_1\,\sigma_2}\left.a_{\sigma\, \sigma'}\right|_{M'}\,C^{s_T\, \sigma_T}_{s_T\, \lp\sigma_T-M'\rp\, L\, M'}.
    }
\end{eqnarray}
Here, $\left.a_{\sigma\, \sigma'}\right|_{\sigma^{-}_T}$ (and $\left.a_{\sigma\, \sigma'}\right|_{M'}$) is $a_{\sigma\, \sigma'}$ after having done the substitutions in $\sigma'$. In particular, for $M'=0$
\begin{eqnarray}
    &\dis{B_{L}=\frac{\sqrt{4\pi}}{a_+}\sum_{\sigma_1\,\sigma_2}\left.a_{\sigma\, \sigma'}\right|_{0}\,C^{s_T\, \sigma_T}_{s_T\, \sigma_T\, L\, 0}=\frac{\sqrt{4\pi}}{a_+}\sum_{\sigma_1\,\sigma_2}a_{\sigma\, \sigma}\,C^{s_T\, \sigma_T}_{s_T\,\sigma_T\, L\, 0}, \hbox{ with }
    }\\ \ecart \nn
    &\dis{a_{\sigma\, \sigma}=\sum_{J\,J'}\lp\dfrac{2J+1}{4\pi}\rp\lp\dfrac{2J'+1}{4\pi}\rp\sum_{\lambda_1}\bra{\sigma_1\,\sigma_2}{S^{J\ \dagger}_{\zeta_\sigma^{-}}}\ket{\lambda_1\,\lp\lambda_1-\zeta_\sigma^{-}\rp}\bra{\lambda_1\,\lp\lambda_1-\zeta_\sigma^{-}\rp}{S^{J'}_{\zeta_\sigma^{-}}}\ket{\sigma_1,\sigma_2}.
    }
\end{eqnarray}
For off-shell initial particles, $\bar{\kappa}=\{\bar{m}_1,\bar{m}_2\}$ and the normalisation factor $\bar{K}$ is added to the $B_{L\,M'}$ expression:
\begin{eqnarray}
    &\dis{B_{L\,M'}=\lp\frac{4\pi}{2L+1}\rp^{1/2} \frac{\tilde{B}_{L\,M'}}{\bar{K}}=\frac{\sqrt{4\pi}}{a_+\bar{K}}\sum_{\sigma_1\,\sigma_2}\left.a_{\sigma\, \sigma'}\right|_{M'}\,C^{s_T\, \sigma_T}_{s_T\, \lp\sigma_T-M'\rp\, L\, M'},
    }\\ \ecart \nn
    &\dis{B_{L}=\frac{\sqrt{4\pi}}{a_+\bar{K}}\sum_{\sigma_1\,\sigma_2}a_{\sigma\, \sigma}\,C^{s_T\, \sigma_T}_{s_T\,\sigma_T\, L\, 0}.
    }
\end{eqnarray}

\subsection{$\bar{2}\to n$}\label{sec:CasesProdMat2n}
The initial and final states have already been studied in previous sections and are given by
\begin{eqnarray}
    \ket{\bar{\mathbb{1}}\, \bar{\kappa}\, \bar{\lambda}}&=&\ket{\bar{\mathbb{1}}\, \bar{\lambda}}\propto\dis\sum_{\bar{J}}\lp 2\bar{J}+1\rp^{1/2}\ket{\bar{J}\,\bar{\zeta}_\lambda^{-}\,\bar{\lambda}},
    \\ \ecart \nn
    \ket{\mathbb{1}\, \kappa\, \lambda}&\propto&\sum_{J}\sum_{\Lambda}\lp 2J+1\rp^{1/2}\ket{J\,\Lambda\,\kappa\,\lambda}.
\end{eqnarray}
Therefore,
\begin{eqnarray}
    &\dis{\Gamma^T_{\bar{\lambda}\, \bar{\lambda}'}\propto\sum_{\lambda}\sum_{J\,J'}\lp 2J+1\rp\,\lp 2J'+1\rp\bra{\bar{\lambda}}{S^{J\ \dagger}_{\bar{\zeta}_\lambda^{-}}}\ket{\kappa\,\lambda}\bra{\kappa\,\lambda}{S^{J'}_{\bar{\zeta}_\lambda^{-\, '}}}\ket{\bar{\lambda}'}\hbox{ and }
    }\\ \ecart \nn
    &\dis{\Gamma^T=\frac{1}{a_{+}}\sum_{\sigma\,\sigma'}a_{\sigma\,\sigma'}\,e_{\sigma\,\sigma'},\hbox{ with }a_{\sigma\, \sigma'}=\sum_{J\,J'}\lp\dfrac{2J+1}{4\pi}\rp\lp\dfrac{2J'+1}{4\pi}\rp\sum_{\lambda}\bra{\sigma}{S^{J\ \dagger}_{\zeta_\sigma^{-}}}\ket{\kappa\,\lambda}\bra{\kappa\,\lambda}{S^{J'}_{\zeta_\sigma^{-\,'}}}\ket{\sigma'}.
    }
\end{eqnarray}
In order to obtain $\tilde{B}_{L\,\sigma^{-}_T}$ and $B_{L\,M'}$, we can reason with the sum over $\lp \sigma,\sigma'\rp$ as in the previous section but only considering the Kronecker delta stemming from $\lp\sigma-\sigma'\rp\cdot d^{(v)}=\sigma_T-\sigma_T'=\sigma_T^{-}$. Nonetheless, due to the new replacements to be done:
\begin{eqnarray}
    &\dis{\sigma_1'=\frac{\sigma_T'+\zeta^{-\,'}_{\sigma}}{d_2+1}=\sigma_1-\frac{1}{d_2+1}\left[\sigma_T^{-}+\lp\zeta^{-}_{\sigma}-\zeta^{-\,'}_{\sigma}\rp\right] \hbox{ and }
    }\\ \ecart \nn
    &\dis{\sigma_2'=\frac{\sigma_T'- d_2\,\zeta^{-\,'}_{\sigma}}{d_2+1}=\sigma_2-\frac{1}{d_2+1}\left[\sigma_T^{-}-d_2\lp\zeta^{-}_{\sigma}-\zeta^{-\,'}_{\sigma}\rp\right],
    }
\end{eqnarray}
the corresponding expressions for the coefficients are not as simple as before. Indeed, we have a sum over $\sigma=\lp\sigma_1,\sigma_2\rp$ and $\zeta^{-\,'}_{\sigma}$:
\begin{eqnarray}
    &\dis{\tilde{B}_{L\,\sigma^{-}_T}=\frac{\lp 2L +1\rp ^{1/2}}{a_{+}}\sum_{\sigma_1\,\sigma_2}\sum_{\zeta^{-\,'}_{\sigma}}\left.a_{\sigma\, \sigma'}\right|_{\sigma^{-}_T}\, C^{s_T\, \sigma_T}_{s_T\, \lp\sigma_T-\sigma_T^{-}\rp\, L\, \sigma_T^{-}},
    }\\ \ecart \nn
    &\dis{B_{L\,M'}=\lp\frac{4\pi}{2L+1}\rp^{1/2}\frac{\tilde{B}_{L\,M'}}{K} =\frac{\sqrt{4\pi}}{a_+ K}\sum_{\sigma_1\,\sigma_2}\sum_{\zeta^{-\,'}_{\sigma}}\left.a_{\sigma\, \sigma'}\right|_{M'}\,C^{s_T\, \sigma_T}_{s_T\, \lp\sigma_T-M'\rp\, L\, M'},
    }
\end{eqnarray}
where $\left.a_{\sigma\, \sigma'}\right|_{\sigma^{-}_T}$ (and $\left.a_{\sigma\, \sigma'}\right|_{M'}$) is $a_{\sigma\, \sigma'}$ after having done the corresponding replacements. The amount of sums can be reduced for certain values of $\sigma_T^{-}$ ($M'$), as it happens for $\sigma_T^{-}=M'=0$ due to the result proven in Appendix \ref{sec:SigmaCond}:
\begin{eqnarray}
    &\dis{B_{L}=\frac{\sqrt{4\pi}}{a_+ K}\sum_{\sigma_1\,\sigma_2}\sum_{\zeta^{-\,'}_{\sigma}}\left.a_{\sigma\, \sigma'}\right|_{0}\,C^{s_T\, \sigma_T}_{s_T\, \sigma_T\, L\, 0}=\frac{\sqrt{4\pi}}{a_+ K}\sum_{\sigma_1\,\sigma_2}a_{\sigma\, \sigma}\,C^{s_T\, \sigma_T}_{s_T\,\sigma_T\, L\, 0}, \hbox{ with }
    }\\ \ecart \nn
    &\dis{a_{\sigma\, \sigma}=\sum_{J\,J'}\lp\dfrac{2J+1}{4\pi}\rp\lp\dfrac{2J'+1}{4\pi}\rp\sum_{\lambda_1}\bra{\sigma_1\,\sigma_2}{S^{J\ \dagger}_{\zeta_\sigma^{-}}}\ket{\lambda_1\,\lp\lambda_1-\zeta_\sigma^{-}\rp}\bra{\lambda_1\,\lp\lambda_1-\zeta_\sigma^{-}\rp}{S^{J'}_{\zeta_\sigma^{-}}}\ket{\sigma_1,\sigma_2}.
    }
\end{eqnarray}
For off-shell initial particles, $\bar{\kappa}=\{\bar{m}_1,\bar{m}_2\}$ and the normalisation factor $\bar{K}$ is added to the $B_{L\,M'}$ expression:
\begin{eqnarray}
    &\dis{B_{L\,M'}=\frac{\sqrt{4\pi}}{a_+\bar{K} K}\sum_{\sigma_1\,\sigma_2}\sum_{\zeta^{-\,'}_{\sigma}}\left.a_{\sigma\, \sigma'}\right|_{M'}\,C^{s_T\, \sigma_T}_{s_T\, \lp\sigma_T-M'\rp\, L\, M'},
    }\\ \ecart \nn
    &\dis{B_{L}=\frac{\sqrt{4\pi}}{a_+\bar{K} K}\sum_{\sigma_1\,\sigma_2}a_{\sigma\, \sigma}\,C^{s_T\, \sigma_T}_{s_T\,\sigma_T\, L\, 0}.
    }
\end{eqnarray}

\subsection{$\bar{m}\to 2$}\label{sec:CasesProdMatm2}
The general roles of the initial and final states are exchanged with respect to the previous scenario:
\begin{eqnarray}
    \ket{\bar{\mathbb{1}}\, \bar{\kappa}\, \bar{\lambda}}&\propto&\dis\sum_{\bar{J}}\sum_{\bar{\Lambda}}\lp 2\bar{J}+1\rp^{1/2}\ket{\bar{J}\,\bar{\Lambda}\,\bar{\kappa}\, \bar{\lambda}},
    \\ \ecart \nn
    \ket{\mathbb{1}\, \kappa\, \lambda}&=&\ket{\mathbb{1}\, \lambda}\propto\sum_{J}\lp 2J+1\rp^{1/2}\ket{J\, \zeta_\lambda^{-}\,\lambda}.
\end{eqnarray}
In consequence, 
\begin{eqnarray}
    &\dis{\Gamma^T_{\bar{\lambda}\, \bar{\lambda}'}\propto\sum_{\lambda_1\, \lambda_2}\sum_{J\,J'}\lp 2J+1\rp\,\lp 2J'+1\rp\bra{\bar{\kappa}\, \bar{\lambda}}{S^{J\ \dagger}_{\zeta_\lambda^{-}}}\ket{\lambda}\bra{\lambda}{S^{J'}_{\zeta_\lambda^{-}}}\ket{\bar{\kappa}\, \bar{\lambda}'}\hbox{ and }
    }\\ \ecart \nn
    &\dis{\Gamma^T=\frac{1}{a_{+}}\sum_{\sigma\,\sigma'}a_{\sigma\,\sigma'}\,e_{\sigma\,\sigma'},\hbox{ with }a_{\sigma\, \sigma'}=\sum_{J\,J'}\lp\dfrac{2J+1}{4\pi}\rp\lp\dfrac{2J'+1}{4\pi}\rp\sum_{\lambda_1\, \lambda_2}\bra{\bar{\kappa}\, \sigma}{S^{J\ \dagger}_{\zeta_\lambda^{-}}}\ket{\lambda}\bra{\lambda}{S^{J'}_{\zeta_\lambda^{-}}}\ket{\bar{\kappa}\, \sigma'}.
    }
\end{eqnarray}
Regarding the expressions of the $\tilde{B}_{L\,\sigma_T^{-}}$, $B_{L\,M'}$ and $B_L$ coefficients, in general the $\sigma_T^{-}$ constraint does not significantly help in simplifying them. However, we recall their definitions:
\begin{eqnarray}
    &\dis{\tilde{B}_{L\,\sigma_T^{-}}=\dfrac{\lp 2L +1\rp ^{1/2}}{a_{+}}\sum_{\substack{\sigma\, \sigma'\\ \lp\sigma-\sigma'\rp\cdot d^{(v)}=\sigma_T^{-}}}\left.a_{\sigma\, \sigma'}\right|_{\sigma^{-}_T}\, C^{s_T\, \sigma_T}_{s_T\, \lp\sigma_T-\sigma_T^{-}\rp\, L\, \sigma_T^{-}},
    }\\ \ecart \nn
    &\dis{B_{L\,M'}=\frac{\sqrt{4\pi}}{a_+\bar{K}}\sum_{\substack{\sigma\, \sigma'\\ \lp\sigma-\sigma'\rp\cdot d^{(v)}=M'}}\,\left.a_{\sigma\, \sigma'}\right|_{M'}\,C^{s_T\, \sigma_T}_{s_T\, \lp\sigma_T-M'\rp\, L\, M'},\quad B_{L}=\frac{\sqrt{4\pi}}{a_+\bar{K}}\sum_{\sigma}a_{\sigma\, \sigma}\, C^{s_T\, \sigma_T}_{s_T\, \sigma_T\, L\, 0},
    }\\ \ecart \nn
    &\hbox{ with }\dis{a_{\sigma\, \sigma}=\sum_{J\,J'}\lp\dfrac{2J+1}{4\pi}\rp\lp\dfrac{2J'+1}{4\pi}\rp\sum_{\lambda_1\, \lambda_2}\bra{\bar{\kappa}\, \sigma}{S^{J\ \dagger}_{\zeta_\lambda^{-}}}\ket{\lambda}\bra{\lambda}{S^{J'}_{\zeta_\lambda^{-}}}\ket{\bar{\kappa}\, \sigma}.
    }
\end{eqnarray}
For off-shell initial particles, the $\bar{\kappa}$ set of parameter increases accordingly. Nevertheless the general form of the $B_{L\,M'}$ coefficient remains invariant.

\subsection{$\bar{m}\to n$}\label{sec:CasesProdMatmn}
For completeness, let us analyse the general scenario, for which
\begin{equation}\left.\begin{array}{l}
    \ket{\bar{\mathbb{1}}\, \bar{\kappa}\, \bar{\lambda}}\propto\dis\sum_{\bar{J}}\sum_{\bar{\Lambda}}\lp 2\bar{J}+1\rp^{1/2}\ket{\bar{J}\,\bar{\Lambda}\,\bar{\kappa}\, \bar{\lambda}}
    \\ \ecart \nn
    \ket{\mathbb{1}\, \kappa\, \lambda}\propto\dis\sum_{J}\sum_{\Lambda}\lp 2J+1\rp^{1/2}\ket{J\,\Lambda\,\kappa\,\lambda}
    \end{array}\right\}\implies
\end{equation}
\begin{eqnarray}
    &\implies\dis{\Gamma^T_{\bar{\lambda}\, \bar{\lambda}'}\propto\sum_{\lambda}\sum_{J\,J'}\lp 2J+1\rp\,\lp 2J'+1\rp\sum_{\Lambda\,\Lambda'}\bra{\bar{\kappa}\, \bar{\lambda}}{S^{J\ \dagger}_{\Lambda}}\ket{\kappa\,\lambda}\bra{\kappa\,\lambda}{S^{J'}_{\Lambda'}}\ket{\bar{\kappa}\, \bar{\lambda}'}\hbox{ and }
    }\\ \ecart \nn
    &\dis{\Gamma^T=\frac{1}{a_{+}}\sum_{\sigma\,\sigma'}a_{\sigma\,\sigma'}\,e_{\sigma\,\sigma'},\hbox{ with }a_{\sigma\, \sigma'}=\sum_{J\,J'}\lp\dfrac{2J+1}{4\pi}\rp\lp\dfrac{2J'+1}{4\pi}\rp\sum_{\Lambda\,\Lambda'}\sum_{\lambda}\bra{\bar{\kappa}\, \sigma}{S^{J\ \dagger}_{\Lambda}}\ket{\kappa\,\lambda}\bra{\kappa\,\lambda}{S^{J'}_{\Lambda'}}\ket{\bar{\kappa}\, \sigma'}.
    }
\end{eqnarray}
Finally, independently of whether the initial particles are on-shell or off-shell, we have 
\begin{eqnarray}
    &\dis{\tilde{B}_{L\,\sigma_T^{-}}=\dfrac{\lp 2L +1\rp ^{1/2}}{a_{+}}\sum_{\substack{\sigma\, \sigma'\\ \lp\sigma-\sigma'\rp\cdot d^{(v)}=\sigma_T^{-}}}\left.a_{\sigma\, \sigma'}\right|_{\sigma^{-}_T}\, C^{s_T\, \sigma_T}_{s_T\, \lp\sigma_T-\sigma_T^{-}\rp\, L\, \sigma_T^{-}},
    }\\ \ecart \nn
    &\dis{B_{L\,M'}=\frac{\sqrt{4\pi}}{a_+ \bar{K} K}\sum_{\substack{\sigma\, \sigma'\\ \lp\sigma-\sigma'\rp\cdot d^{(v)}=\sigma_T^{-}}}\left.a_{\sigma\, \sigma'}\right|_{M'}\,C^{s_T\, \sigma_T}_{s_T\, \lp\sigma_T-M'\rp\, L\, M'},\quad B_{L}=\frac{\sqrt{4\pi}}{a_+\bar{K} K}\sum_{\sigma}a_{\sigma\, \sigma}\, C^{s_T\, \sigma_T}_{s_T\, \sigma_T\, L\, 0},
    }\\ \ecart \nn
    &\hbox{ with }\dis{a_{\sigma\, \sigma}=\sum_{J\,J'}\lp\dfrac{2J+1}{4\pi}\rp\lp\dfrac{2J'+1}{4\pi}\rp\sum_{\Lambda\,\Lambda'}\sum_{\lambda}\bra{\bar{\kappa}\, \sigma}{S^{J\ \dagger}_{\Lambda}}\ket{\kappa\,\lambda}\bra{\kappa\,\lambda}{S^{J'}_{\Lambda'}}\ket{\bar{\kappa}\, \sigma}.
    }
\end{eqnarray}

\section{Proof for the $\sigma_T^{-}=0$ condition}\label{sec:SigmaCond}
We want to prove that $\sigma^{-}_T=0$ is equivalent to $\sigma=\sigma'$. From the definition of $\sigma_T^{-}$ the necessary condition is trivial, so let us focus on the sufficient one. Because of $d^{(v)}\neq0$, 
\begin{equation}
    \sigma_T^{-}=0 \iff \lp \sigma-\sigma'\rp\cdot d^{(v)}=0\iff \sigma=\sigma' \vee \lp \sigma-\sigma'\rp \perp d^{(v)}.
\end{equation}
We proceed by induction in $m$: \medskip

\noindent
\underline{\textbf{Case $m=2$}}\\ 
Let us assume that $\sigma\neq\sigma'$, we have
\begin{eqnarray}
   \sigma_T^{-}=0\iff d_2 \lp \sigma_1-\sigma'_1\rp+\lp \sigma_2-\sigma'_2\rp=0\iff d_2 \lp \sigma_1-\sigma'_1\rp=-\lp \sigma_2-\sigma'_2\rp
\end{eqnarray}

Taking into account the previous relation, $\sigma\neq\sigma'\iff\lp \sigma_1-\sigma'_1\rp,\lp \sigma_2-\sigma'_2\rp\neq0$. Moreover, due to $\sigma_1,\sigma_1'\in\left\{-\lp\frac{d_1-1}{2}\rp,\dots,\frac{d_1-1}{2}\right\}$, the combination $\lp \sigma_1-\sigma'_1\rp$ is a non-zero integer that takes values $\lp \sigma_1-\sigma'_1\rp\in\left\{-\lp d_1-1\rp,\dots,\lp d_1-1\rp\right\}\setminus\{0\}$ and same with $\lp \sigma_2-\sigma'_2\rp$ exchanging $d_1\to d_2$. Hence,
\begin{eqnarray}
    d_2\leq |d_2 \lp \sigma_1-\sigma'_1\rp|=|\sigma_2-\sigma'_2|\leq d_2-1,
\end{eqnarray}
leading to a contradiction. Therefore $\sigma=\sigma'$.
\medskip

\noindent
\underline{\textbf{Case $m$}}\\ 
Assuming the result to hold for $m-1$, let us see it for $m$. In this case, if $\sigma\neq\sigma'$
\begin{eqnarray}
   \sigma_T^{-}=0\iff-\lp \sigma_m-\sigma'_m\rp=\sum_{i=1}^{m-1} \lp \sigma_i-\sigma'_i\rp d^{(v)}_i=d_m \sum_{i=1}^{m-1} \lp \sigma_i-\sigma'_i\rp \left.d^{(v)}_i\right|_{m-1}=d_m \left.\sigma_T^{-}\right|_{m-1},
\end{eqnarray}
where we have identified the last two terms in the previous equation with the $ d^{(v)}$ and $\sigma_T^{-}$ associated with the $m-1$ case. 

By inspection of the previous relation, $\sigma_m-\sigma'_m=0\iff \left.\sigma_T^{-}\right|_{m-1}=0$, so if $\sigma_m-\sigma'_m=0$ the induction hypothesis applied in $\left.\sigma_T^{-}\right|_{m-1}=0$ guarantees $\sigma=\sigma'$. For $\sigma_m-\sigma'_m\neq0$, we follow the same reasoning than in the $m=2$ case. We know that $\lp \sigma_m-\sigma'_m\rp\in\left\{-\lp d_m-1\rp,\dots,\lp d_m-1\rp\right\}\setminus\{0\}$ and $\left.\sigma_T^{-}\right|_{m-1}\in\left\{-\lp d/d_m-1\rp,\dots,\lp d/d_m-1\rp\right\}\setminus\{0\}$. Thus,
\begin{eqnarray}
    d_m\leq |d_m\left.\sigma_T^{-}\right|_{m-1}|=| \sigma_m-\sigma'_m|\leq d_m-1,
\end{eqnarray}
leading to a contradiction. Therefore, $\sigma=\sigma'$.\cqfd
\newpage

\bibliographystyle{style.bst}  % (uses file "style.bst"  Includes doi, etc)
\bibliography{biblio}	 % references a file(s) called  *.bib -DO NOT INCLUDE EXTENSION - 

\end{document}